\documentclass[twocolumn,superscriptaddress,amsmath,amssymb,floatfix,aps,prl]{revtex4}
\usepackage{adjustbox}
\usepackage{graphicx}
\usepackage{dcolumn}
\usepackage{bm}
\usepackage{color}
\usepackage{hyphenat}
\usepackage[normalem]{ulem}
\usepackage{soul}
\usepackage{hyperref}

\newcommand{\fakesection}[1]{%
  \par\refstepcounter{section}
  \sectionmark{#1}
  \addcontentsline{toc}{section}{\protect\numberline{\thesection}#1}
}

\begin{document}
 \raggedbottom

\title{The behaviour of information flow near criticality}

\author{Matthijs Meijers}
\affiliation{NWO institute AMOLF, Science Park 104, 1098 XG Amsterdam, The Netherlands }
\author{Sosuke Ito}
\affiliation{NWO institute AMOLF, Science Park 104, 1098 XG Amsterdam, The Netherlands }
\affiliation{Universal Biology Institute, The University of Tokyo,
  7-3-1 Hongo, Bunkyo-ku, Tokyo, 113-0033, Japan}
\author{Pieter Rein ten Wolde}%
\affiliation{%
 NWO institute AMOLF, Science Park 104, 1098 XG Amsterdam, The Netherlands }%

\date{\today}

\begin{abstract}
  Recent experiments have indicated that many biological
  systems self-organise near their critical point, which hints at a
  common design principle. While it has been suggested that
  information transmission is optimized near the critical point, it
  remains unclear how information transmission depends on the dynamics
  of the input signal, the distance over which the information needs
  to be transmitted, and the distance to the critical
  point.  Here we employ stochastic simulations of a driven 2D Ising
  system and study the instantaneous mutual information and the
  information transmission rate between a driven input spin
  and an output spin.  The instantaneous mutual information varies
  non-monotonically with the temperature, but increases monotonically
  with the correlation time of the input signal.  In contrast, the
  information transmission rate exhibits a maximum as a function of
  the input correlation time. Moreover, there exists an optimal
  temperature that maximizes this maximum information transmission
  rate. It arises from a tradeoff between the necessity to respond
  fast to changes in the input so that more information per unit
  amount of time can be transmitted, and the need to respond to
  reliably. The optimal temperature lies above the critical point, but
  moves towards it as the distance between the input and output spin
  is increased. 
%
%
\end{abstract}

\maketitle

Most, if not all, living organisms need to respond to changes in their
environment. Examples include bacteria searching for food, animals
trying to catch prey, or birds in flocks trying to coordinate their
motion. In all these cases, the flow of information, be it via an
intracellular biochemical network, an intercellular neural network, or
between the individuals within the group, is vital to
function. Moreover, in all these examples not only the reliability of
information transmission is important, but also the speed and the
distance over which the information needs to be transmitted.

Recent experiments indicate that many biological systems self-organise
at their critical point. Examples are the flocking behaviour of
starlings \cite{Cavagna2010}, signal percolation within a bacterial
community \cite{Larkin:2018ky}, neural networks \cite{Tkacik2014}, and
cooperative transport in ants \cite{Feinerman2018}. The critical point
is a special point in the phase diagram, right at
the border between order and disorder and characterized by the
divergence of thermodynamic variables \cite{Chandler1987}. The
observation that different systems self-organize near the critical
point hints at a common design principle \cite{Mora2011}. In this
letter, we investigate whether criticality enhances the speed and
reliability of information transmission.

The observation that very different systems self-organize near the
critical point makes the Ising system, because of its generic
properties, a natural choice to study information flow. Different
measures have been employed to characterize information transmission
in the Ising system. One is the average mutual information
\cite{Shannon1948} between pairs of neighboring spins
\cite{Matsuda1996,Gu2008}. It quantifies the correlations between
these spins and can thus be interpreted as a measure for the
reliability of information transmission. It exhibits a sharp peak at
the critical point \cite{Matsuda1996,Gu2008}.
However, the mutual information does not contain dynamical aspects of
information flow.  An information theoretic measure that does include
this is the transfer entropy \cite{Schreiber2000}. Barnett \textit{et
  al.} showed that the transfer entropy between pairs of
  neighboring spins peaks at the critical point while a global
  transfer entropy measure peaks above it
\cite{Barnett2013}. However, the authors studied systems in
  thermodynamic equilibrium, while biological systems, if not any
  information-processing device, are driven out of
  equilibrium via the driving of the input signal. Moreover, they
  only studied the one-step transfer entropy, which does not take into
  account that the dynamics of the output may depend on the history of
  the input.

\begin{figure}[t]
 \centering
 \begin{adjustbox}{center}
   \includegraphics[height=50mm,width=0.8\columnwidth]{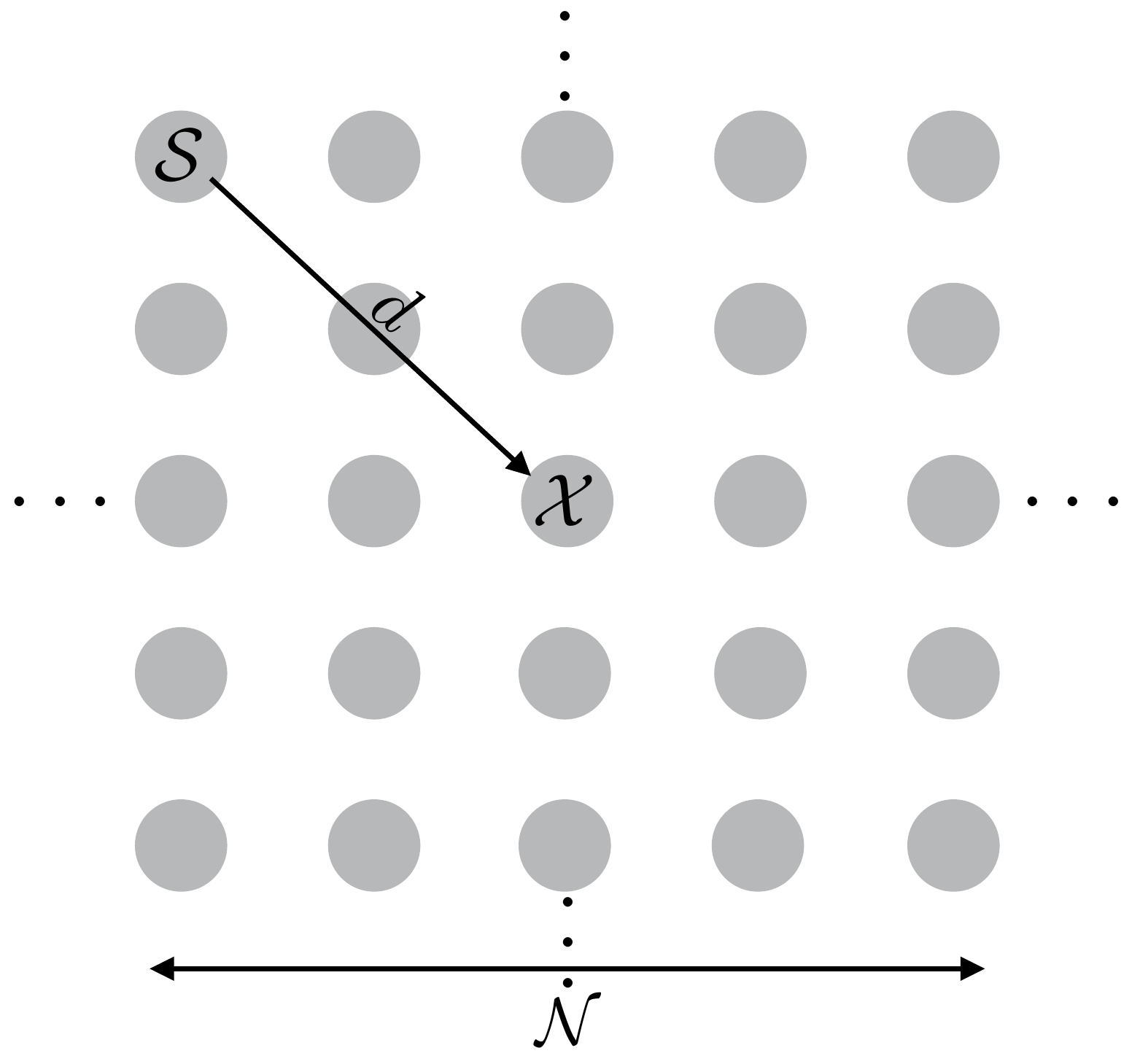}
 \end{adjustbox}
 \caption{We consider an Ising system containing $\mathcal{N} \times
   \mathcal{N}$ spins with periodic boundary conditions. One spin is
   chosen to be the input spin and is flipped according to a
   stationary random-telegraph progress. We measure the
   information transmission from the input to the output spin,
   positioned a distance $d$ along the diagonal from the input;
     $d$ in units of the distance between two neighboring spins along
     the diagonal.}
 \label{SetUp}
\end{figure}

  In this Letter, we investigate the flow of information in a
  driven 2D Ising system. The input signal $\mathcal{S}$ is a
  spin, the {\em input spin}, which is flipped according to a
  stationary random telegraph process with a time scale
  $\tau_{\rm s}$. Since this driving signal is stationary, our system
  is in a stationary non-equilibrium steady-state.
 The output signal $\mathcal{X}$ is another spin,
  {\em the output spin}, which is located at a distance $d$ away from
  the input spin, see Fig.~\ref{SetUp}. The input and output signals
  produce the random variables $S,X \in \{+1, -1\}$ at each point in
  time, respectively.  Because the information
  is propagated from the input  to the output spin via the other
  spins, the dynamics of the output are distinctly
  non-Markovian. Consequently, we need to recognize the history of the
  input and output signal in characterizing the information flow
  between them.

To characterize information transmission, we study two measures: the
{\em instantaneous} mutual information $I_{\rm inst}$ and the
information transmission rate $I_{\rm R}$. The measure
$I_{\rm inst}(S;X)$ is the mutual information between the
stationary input and
output signal at a single point in time:
\begin{equation}
I_{\rm inst}(S;X) = H(S) - H(S|X),
\end{equation}
where $H(S)$ is the Shannon entropy of the input signal and $H(S|X)$
is the Shannon entropy conditional on the output signal. The
instantaneous mutual information has been used to study information
transmission in intracellular signaling networks
\cite{Tostevin2010,Brittain:2017hf,Das:2017gh,Malaguti:2019ww}. In
contrast to the mutual information studied by Matsuda et
al. \cite{Matsuda1996}, which characterizes equilibrium correlations
between pairs of spins, $I_{\rm inst}(S;X)$ quantifies the
non-equilibrium correlations between the input and output spin of our
driven system.  These correlations depend on the input time scale
$\tau_{\rm s}$ and the response time $\tau_{\rm r}$ of the system,
which is determined by the temperature. 

While the instantaneous mutual information $I_{\rm inst}(S;X)$ quantifies
how accurately the input spin is mapped onto the output spin, it does
not quantify the rate of information transmission. The latter is not
only determined by the accuracy of the input-output mapping, but also
by the rate at which independent ``messages'' are transmitted through
the system. Autocorrelations within the  input and the output signal 
lower the information transmission rate.
  To take these correlations into
account, we study the information transmission rate, which  is
defined as the rate at which the mutual information between the
trajectories of the input and output signal increases
\cite{Tostevin2009}:
\begin{equation}
I_{\text{R}} = \frac{1}{L} \lim_{L \rightarrow \infty} I(\mathbf{S}_{L};\mathbf{X}_{L}),
\label{eq:IR}
\end{equation}
where $ \mathbf{S}_{L}=[S(t_1), S(t_{2}), \dots, S(t_n)] $ and
$\mathbf{X}_{L}=[X(t_1), X(t_{2}), \dots, X(t_n)]$ are spin
trajectories of duration $L=(n-1)\delta t$, containing $n$ subsequent
spin states $S$ ($X$) at successive time points $t_i= (i-1) \delta t$,
with $\delta t$ the elementary timestep of the dynamics \cite{SI}. To capture
the autocorrelations in the input and output signal, the trajectory
lengths have to be longer than the longest timescale in the problem,
$L> \tau_{\rm s}, \tau_{\rm r}$; $I_{\rm R}$ then properly takes
into account the history of the input and output spin, in contrast to
the one-step transfer entropy. We note
that 
if
there were no autocorrelations in the input and output signal, the
information transmission rate would reduce to $I_{\rm R} = I_{\rm
  inst} / \delta t$. However, in general, $I_{\rm R}$ is lower than
  $I_{\rm inst} / \delta t$, precisely because of the signal
  autocorrelations. We also point out that since the output signal
  does not feed back on the input, the information transmission
  rate can be related to the multi-step transfer entropy
\cite{Massey:1990vy}.

In order to evaluate the effects of the dynamics and criticality on
information flow, we will study both measures as a function of the
time scale of the input signal $\tau_{\rm s}$ for different
temperatures $T$ close to the critical temperature $T_{\rm c}$ and
for different distances $d$ between the input and output signal. We
are mainly interested in temperatures higher than the critical
temperature, since for lower temperatures the system freezes down in
the ferromagnetic phase, drastically slowing down information
transmission.  We will show that the non-trivial interaction
between the diverging correlation length and the diverging response
time near the critical point causes the information flow to be optimal
close to, but not at, the critical point. The optimal temperature is
determined by the distance over which the information needs to be
transmitted and the size of the system.

Consider a 2D Ising system of ${\mathcal{N}}
\times {\mathcal{N}}$ spins with periodic boundary conditions and no
external magnetic field. For a spin configuration $\boldsymbol{\sigma}
= \sigma_1, \dots, \sigma_{\mathcal{N}} \in \{+1,-1 \}$, the
Hamiltonian of the system is $H(\boldsymbol{\sigma}) = - J
\sum_{\langle i, j \rangle} \sigma_i \sigma_j$, where $J$ is the
coupling parameter and the sum is taken over all nearest
neighbours. For isotropic coupling, the critical temperature is
$k_{\rm B}T_{\rm c} / J = 2.269$ \cite{Onsager1944}. Following Barnett
\textit{et al.} \cite{Barnett2013}, we use discrete-time Glauber
spin-flip dynamics \cite{Glauber1963}. We define the response time
$\tau_{\rm r}$  as the time scale over which spontaneous fluctuations in the
  undriven system, as computed via the two-point time correlation
  function of the input and output spin, relax to equilibrium
  \cite{Chandler1987}. 
 Entropies are measured in nats.

The information transmission rate is notoriously difficult to compute,
because the state space of the input and output trajectories rapidly
diverges with the length of the trajectories. We have therefore
considered not only relatively small systems, but also developed
the following scheme: To limit the size of the state space, we
introduce a sampling interval $\Delta t$ such that the trajectory
length $L=(n-1)\Delta t$, where $n$ is the number of spin states in
both the input and output trajectory. As described in \cite{SI}, we
verify that $L$ is longer than the input and output correlation time
such that $I(\mathbf{S}_L;\mathbf{X}_L)$ increases linearly with $L$
and the information transmission rate $I_{\rm R}$ is independent of
$L$. We then compute for long enough $L$, $I_{\rm R}=I_{\rm R}(\Delta t$) for a
range of $\Delta t$ values, where we verify that the entropy
histograms are sampled accurately, using the Bayesian entropy
estimator of Nemenman {\it et al.} to enhance the estimate of the
(joint) entropies \cite{Nemenman2004}. We then extrapolate $I_{\rm R}(\Delta
t)$ to the quantify of interest, $I_{\rm R}(\Delta t\to \delta t)$, where
$\delta t$ is the elementary time step of the Glauber dynamics; to
verify this extrapolation procedure, we have also recomputed by
simulations $I_{\rm R}(\Delta t)$ for a number of extrapolated $\Delta t$
values (see \cite{SI}).

\begin{figure}[t]
 \centering
 \begin{adjustbox}{center}
   \includegraphics[height=50mm,width=\columnwidth]{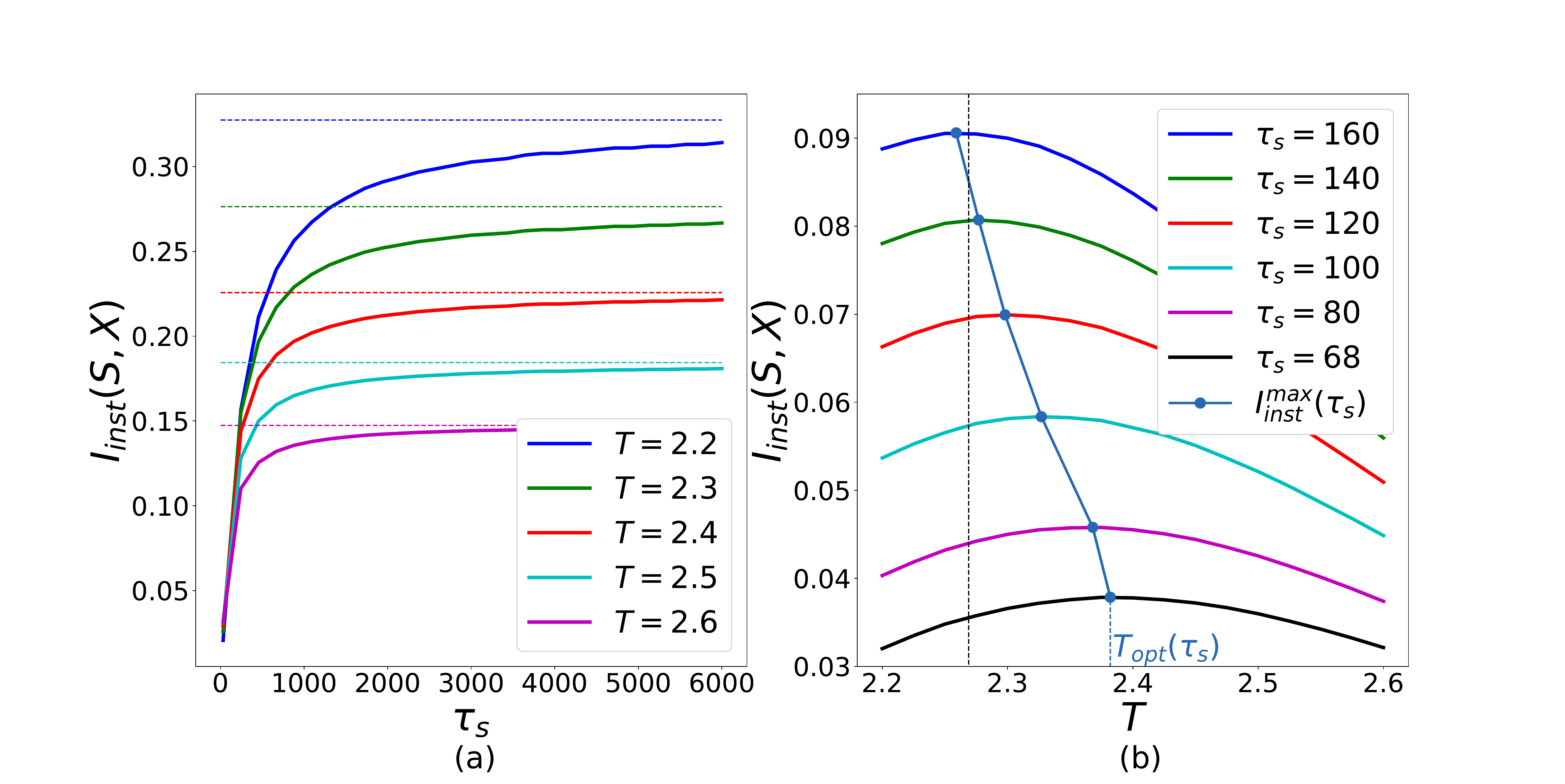}
 \end{adjustbox}
 \caption{The instantaneous mutual information $I_{\rm inst}(S;X)$ as
   a function of the input correlation time $\tau_{\rm s}$ and
   temperature $T$. (a) $I_{\rm inst}(S;X)$ increases monotonically
   with the correlation time of the input signal $\tau_{\rm s}$ until
   it reaches a plateau, $I_{{\rm inst}, \infty}(S;X)$, which is equal
   to the static mutual information (dashed line).  The plateau value
   $I_{{\rm inst}, \infty}(S;X)$ increases as the temperature $T$
   decreases. However, for small $\tau_{\rm s}$, when $\tau_{\rm s}$ is on the
   order of the response time $\tau_{\rm r}$, $I_{\rm inst}(S;X)$ does
   not rise monotonically with decreasing temperature. This is more
   clearly illustrated in panel b, which shows $I_{\rm inst}(S;X)$ as
   a function of the temperature $T$ for different input correlation
   times $\tau_{\rm s}$. It is seen that there exist an optimal
   temperature ${T}_{\rm opt}$ that maximises $I_{\rm inst}(S;X)$ for
   a given $\tau_{\rm s}$; moreover, ${T}_{\rm opt}$ decreases when
   $\tau_{\rm s}$ increases. The optimal temperature arises because a
   higher temperature allows the system to respond more rapidly to
   changes in the input, yet also makes this response more
   noisy. Vertical dashed line denotes critical temperature. The size
   of the system is $5\times 5$, and the distance between input and
   output spin is $d=2$. }
 \label{InstMI}
\end{figure}

Figure~\ref{InstMI} shows the instantaneous mutual information $I_{\rm
  inst}(S;X)$ between the input and output signal separated by a
distance $d=2$ as a function of the input correlation time $\tau_{\rm
  s}$ and temperature $T$ in an Ising system of 5 $\times$ 5
spins. The instantaneous mutual information rises with the input
correlation time $\tau_{\rm s}$ (Fig. \ref{InstMI}a), because this
gives the system more time to respond to changes in the input signal
and hence more time to correlate the output with the input signal. For
large $\tau_{\rm s}$, the instantaneous mutual information reaches a
plateau value $I_{{\rm inst},\infty}(S;X)$ that corresponds to the
static mutual information, which is the mutual information
between the output spin and the input spin when the latter is held
fixed indefinitely for each realization $S=1,-1$. The static mutual
information increases as the temperature is decreased, because
decreasing the temperature lowers the thermal noise in the transmitted
signal.

Panel b of Fig.~\ref{InstMI} shows that for a given correlation time
$\tau_{\rm s}$ of the input signal, there exists an optimal
temperature $T_{\rm opt}$ that maximises the instantaneous mutual
information $I_{\rm inst}(S;X)$. Increasing the temperature raises the
thermal noise in the signal, which tends to lower the instantaneous
mutual information. On the other hand, increasing the temperature also
reduces the response time $\tau_{\rm r}$. This allowes the system to
more accurately track the input signal, which tends to raise the
instantaneous mutual information between the input and output
signal. The interplay between these two effects gives rise to an
optimal temperature $T_{\rm opt}(\tau_{\rm s})$ that maximizes the
instantaneous mutual information, $I_{\rm inst}^{\rm max}(\tau_{\rm
  s})$. This optimal temperature decreases as the input correlation
time $\tau_{\rm s}$ is increased, because this gives the system more
time to respond to the variations in the input. Moreover, the
maximum instantaneous mutual information $I_{\rm inst}^{\rm
  max}(\tau_\text{s})$ rises with $\tau_{\rm s}$, not only because
  increasing $\tau_{\rm s}$ raises $I_{\rm inst}$ by itself, but also
  because the lower optimal temperature $T_{\rm opt}(\tau_{\rm s})$
  reduces the thermal noise in the signal.

\begin{figure}[t]
 \centering
 \begin{adjustbox}{center}
   \includegraphics[height=50mm,width=\columnwidth]{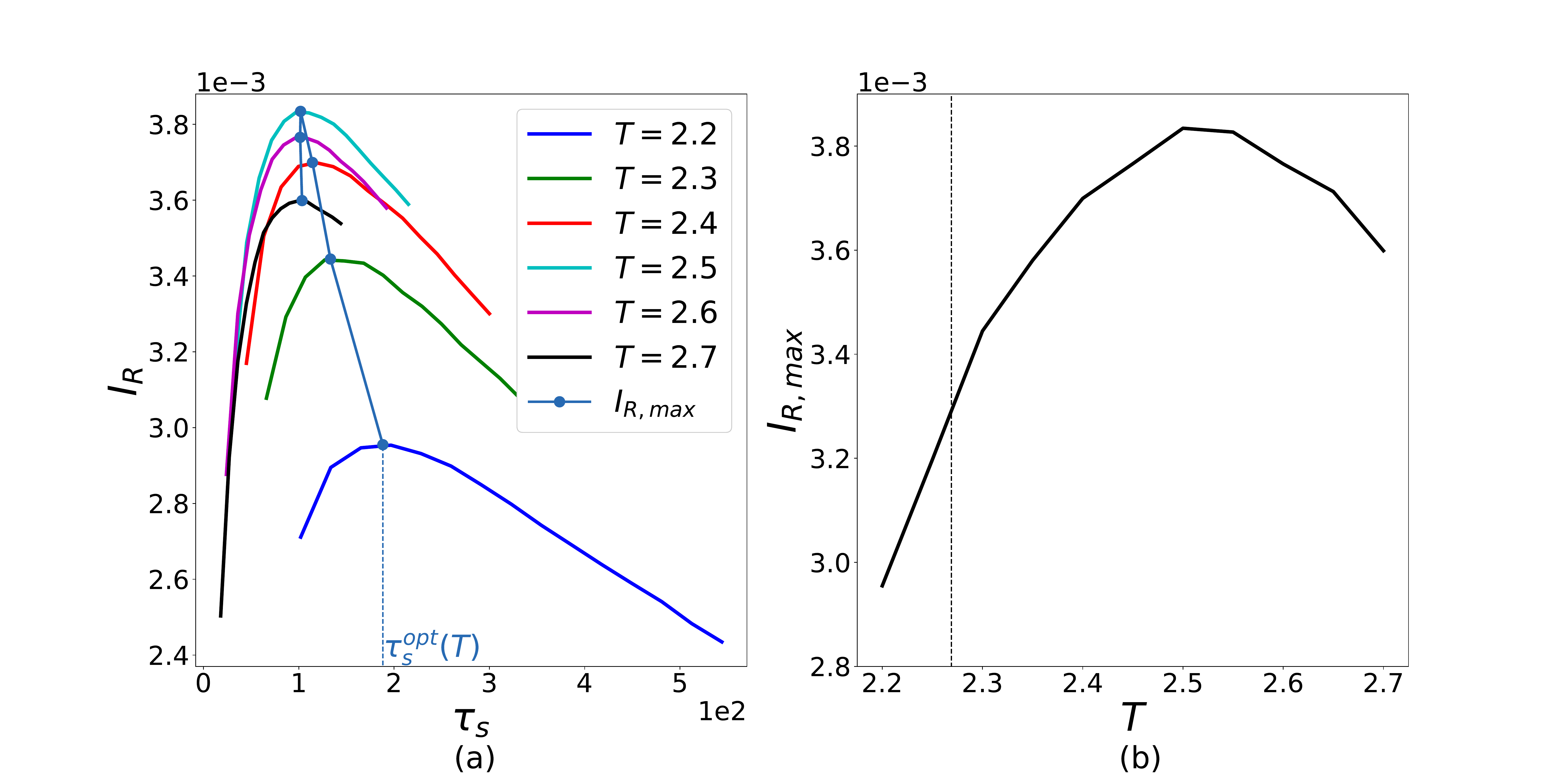}
 \end{adjustbox}
 \caption{The information transmission rate $I_{\rm R}$ as a function
   of the input correlation time $\tau_s$ and temperature $T$. (a) For
   a given temperature ${T}$, there exists an optimal $\tau_{\rm s}$
   that maximises the information transmission rate $I_{\rm
     R}$. Increasing $\tau_{\rm s}$ gives the system more time to
   respond to changes in the input, which enhances the reliability of
   the response and thus lowers $H(\textbf{S}_L|\textbf{X}_L)$ (see
   also Fig.~\ref{InstMI}). Yet, increasing $\tau_{\rm s}$ also
   decreases the number of distinct input states that are transmitted
   per unit amount of time, which reduces the entropy of the input signal
   $H(\textbf{S}_L)$ (see Eq.~\ref{eq:IR}). This interplay gives rises
   to an optimal $\tau_{\rm s}$ at which $I_{\rm R}$ reaches its
   maximal value $I_{\rm R,max}$. The figure also shows that $I_{\rm
     R,max}$ initially rises with $T$, but then decreases, which is
   more clearly illustrated in panel b: there exists an optimal
   temperature $T_{\rm opt}$ that maximises the information
   transmission rate. Vertical dashed line
   denotes critical temperature. System size is $5\times 5$, and
   the distance between input and output spin is $d=2$. Lines are
   truncated at high $\tau_{\rm s}$ for large $T$, because it becomes
   exceedingly difficult to get good statistics in this regime.}

 \label{IR}
\end{figure}

Figure \ref{IR}a shows the information transmission rate $I_{\rm R}$
as a function of the correlation time of the input signal $\tau_{\rm
  s}$ for different temperatures $T$. While, for a given temperature,
the instantaneous mutual information $I_{\rm inst}$ increases
monotonically with the input correlation time $\tau_{\rm s}$ (see
Fig. \ref{InstMI}a), the information transmission rate $I_{\rm R}$
exhibits an optimal $\tau_{\rm s}$ that maximizes the information
transmission rate.  When $\tau_{\rm s}$ is too short, the signal is
changing faster than the output can respond to, which tends to
decrease $I_{\rm R}$ by increasing the conditional entropy
$H(\textbf{S}_L|\textbf{X}_L)$ (see Eq.~\ref{eq:IR}). On the other
hand, for large $\tau_{\rm s}$ time is wasted when the output has been
correlated to the input yet is waiting for the signal to change again;
indeed, the entropy of the input signal $H(\textbf{S}_L)$ rises as
$\tau_{\rm s}$ is decreased, which tends to enhance $I_{\rm R}$ (see
Eq.~\ref{eq:IR}). This interplay causes the information transmission
rate to have a maximum at an optimal input time scale $\tau_{\rm
  s}^{\rm opt}$.
The value of $\tau_{\rm s}^{\rm opt}$ decreases with temperature,
because at higher temperatures the system can respond more rapidly to
changes in the input signal.

Panel b of Fig. \ref{IR} shows the maximum value of the information
transmission rate $I_{\rm R}$ at the optimal input correlation time
$\tau_{\rm s}$, $I_{\rm R, max}$, as a function of the temperature
$T$. Clearly, there exists an optimal temperature that maximizes
$I_{\rm R,max}$. This is in marked contrast to the maximum value of
the instantaneous mutual information, obtained for $\tau_{\rm s} \to
\infty$, which increases monotonically with decreasing temperature,
even for temperatures below $T_{\rm c}$ (see Fig.~\ref{InstMI}a).  The
optimum in $I_{\rm R,max}$ arises from the trade-off between a faster
response at higher temperatures, which allows for a more rapidly
varying input, thereby increasing the entropy of the input signal (see
Eq. 1), and less thermal noise in the transmitted signal at lower
temperatures. That $I_{\rm R,max}$ peaks above the critical
temperature is because the response time rapidly increases near the
critical temperature, thereby decreasing the amount of information per
unit time that can be sent through the system.

\begin{figure}[t]
 \centering
 \begin{adjustbox}{center}
   \includegraphics[height=60mm,width=0.9\columnwidth]{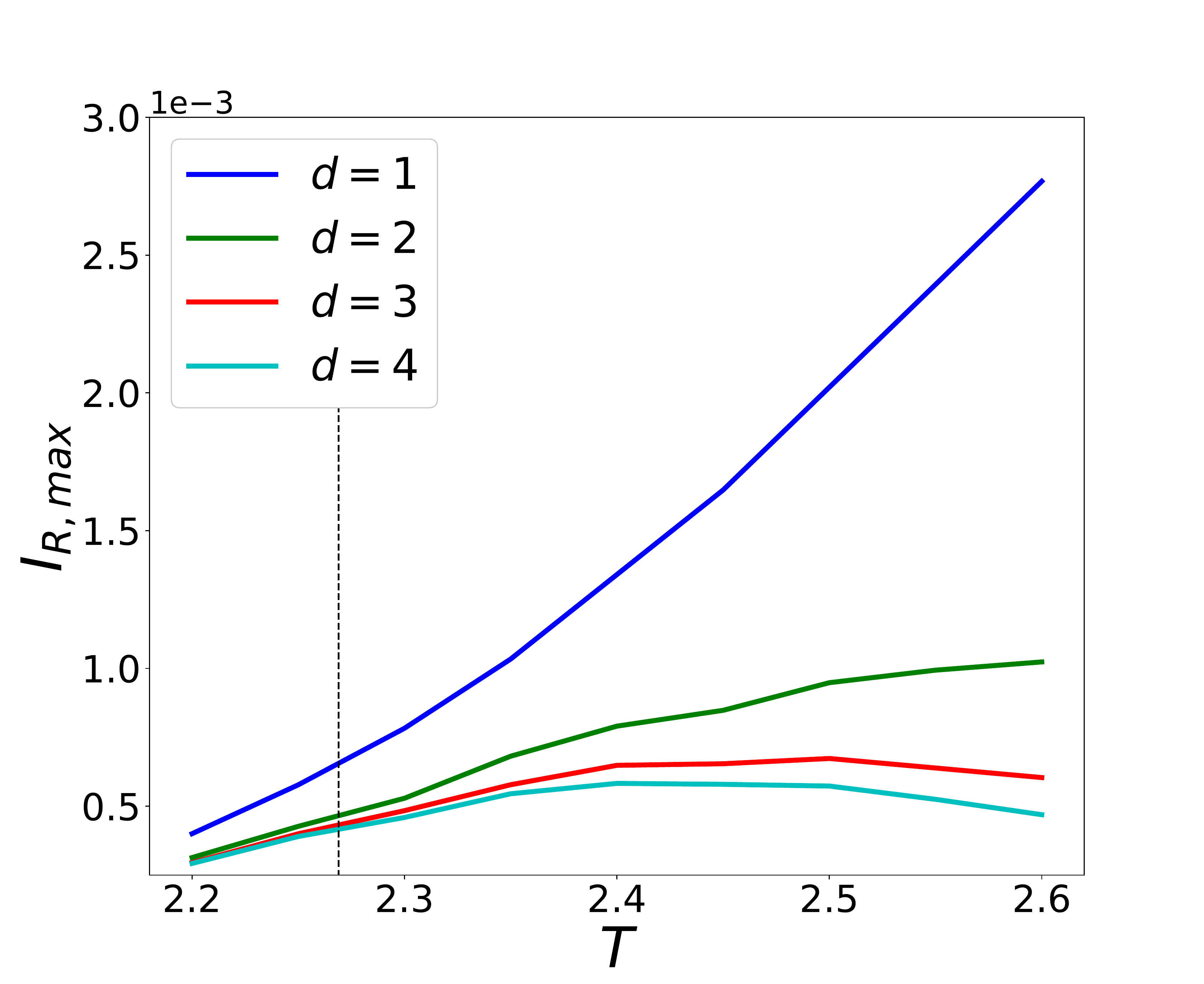}
 \end{adjustbox}
 \caption{The optimal temperature that maximises the information
   transmission rate $I_{\rm R}$ decreases as the distance $d$ over
   which the information is transmitted increases. The figure shows
   for different values of $d$ the maximum value of $I_{\rm R}$,
   $I_{\rm R,max}$, obtained by optimizing $I_{\rm R}$ over the input
   correlation time $\tau_{\rm s}$ (see Fig.~\ref{IR}a), as a
   function of the temperature ${T}$. It is seen that $I_{\rm R,max}$
   decreases as $d$ is increased, while the optimal temperature
   moves closer to the critical temperature, denoted by vertical
   dashed line. The system size is $10\times 10$. }
 \label{MaxIR}
\end{figure}

So far we have kept both the distance $d$ between the input and output
spin constant, as well as the system size. We now systematically vary
these parameters. Figure \ref{MaxIR} shows the maximum information
transmission rate $I_{\rm R,max}$, obtained by optimizing over
  $\tau_{\rm s}$ (see Fig.~\ref{IR}) , as a function of
temperature $T$ for different values of $d$ in a $10 \times 10$ Ising
system. The information transmission rate decreases as $d$ is
increased, because the correlations between spins become weaker as the
distance between them becomes larger.
More interestingly, the optimal temperature that maximizes $I_{\rm
  R,max}$ moves closer to the critical temperature when the distance
between the signal and the output is increased.  When the distance
between the input and output spin is increased, the correlation length
must be increased in order to maintain the correlations between
the input and output spin. This can be achieved by bringing the system
closer to the critical point.


 Critical effects are stronger in systems of larger size.  Close to
 $T_{\rm c}$, the response time of our system increases up to six-fold
 when the system size is increased from $5 \times 5$ to $10 \times 10$
 spins. This makes it beneficial for information
 transmission to move the system further away from the critical point
 when the system size is increased at constant $d$. Compare the case
 of $d=2$ in the $5 \times 5$ system in Fig. \ref{IR} with $d=2$ in
 the $10 \times 10$ system in Fig. \ref{MaxIR}: the optimal
 temperature shifts from ${T}_{\rm opt} \approx 2.5$ to $2.6$ in the
 larger system. The maximum information transmission rate
 $I_{\rm R,max}$ decreases because of the larger response time.

 As the system is moved closer to the critical temperature, both the
 correlation length and the correlation time increase, which have
 opposite effects on information transmission. Moreover, these effects
 increase with the system size, diverging in the thermodynamic
 limit. Hence, when $d$ is increased in a system of constant
 size, the optimal temperature $T_{\rm opt}$ that maximizes
 information transmission moves towards $T_{\rm c}$, because
 the increasing correlation length is necessary to transmit
 information over large distances, as Fig. \ref{MaxIR} shows. It also
 means that when the system size is increased at constant $d$, $T_{\rm
   opt}$ moves away from $T_{\rm c}$, because of the increasing
 response time, as discussed above. This raises the question how
 $T_{\rm opt}$ changes as $d$ is scaled {\em together} with the system
 size, which, as renormalization group theory indicates,
 is also the relevant finite-size scaling question for this problem.
 We have therefore also performed simulations for $d=6$ and ${\cal
   N}=15$. The optimal temperature that maximizes information
 transmission decreases from $T_{\rm opt} \approx 2.53$ for ($d=2$,
 ${\cal N}=5$), to $T_{\rm opt} \approx 2.44$ for ($d=4$, ${\cal
   N}=10$), and $T_{\rm opt}\approx 2.38$ for ($d=6$, ${\cal N}=15$)
 (see Fig. S9 \cite{SI}). Our results thus suggest that $T_{\rm opt}$
 moves towards $T_{\rm c}$ in the thermodynamic limit.

 In summary, the information transmission rate is a dynamic quantity
 that is influenced by both the strength of the correlation that can
 be achieved between the input and the output signal, and the time
 scale on which the output can respond to a change in the input
 signal. These two properties are oppositely influenced by the
 temperature of the system. The system faces a trade-off between
 increasing the temperature in order to decrease the response time
 such that more information per unit time can be transmitted through
 the system and decreasing the temperature in order to increase the
 correlations between the input signal and output. This trade-off
 produces a maximum information transmission rate at an optimal
 temperature that depends on the distance between the in- and output
 spin and the size of the system. The optimal temperature is close to
 yet above the critical point, although our results leave open the
 possibility that it moves towards the critical temperature in the
 thermodynamic limit. Our results may explain why a number of
 biological systems appear to be tuned near the critical point
 \cite{Cavagna2010,Larkin:2018ky,Tkacik2014,Feinerman2018}, and may
 also be relevant for understanding information transfer in systems
 outside the realm of biology. Lastly, many systems, including
 biological systems, are higher dimensional. Since the response time
 does not depend on the dimensionality of the system while
 correlations decay faster with distance in higher dimension, we
 conjecture that in higher dimensional systems the optimal temperature
 is closer to the critical point.

This work is part of the research programme of the
Netherlands Organisation for Scientific Research (NWO)
and was performed at AMOLF. Ito was funded by grants  KAKENHI Grant No. JP16K17780, 
and JST Presto Grant No. JP18070368, Japan. We thank
Ilya Nemenman for sharing his code to estimate entropies and Tom
Ouldridge for a careful reading of our manuscript.



\widetext
\pagebreak
\begin{center}
\textbf{\large Supplement for `The behaviour of information flow near criticality'}
\end{center}

\vspace{0.5cm}

\setcounter{equation}{0}
\setcounter{figure}{0}
\setcounter{table}{0}
\setcounter{page}{1}
\setcounter{section}{0}
\makeatletter
\renewcommand{\theequation}{S\arabic{equation}}
\renewcommand{\thesection}{S\Roman{section}}
\renewcommand{\thefigure}{S\arabic{figure}}
\renewcommand{\bibnumfmt}[1]{[S#1]}
\renewcommand{\citenumfont}[1]{S#1]}

\fakesection{Supplementary Information}
 
The information transmission rate is defined as
\begin{equation}
I_{\text{R}} = \frac{1}{L} \lim_{L \rightarrow \infty} I(\mathbf{S}_{L};\mathbf{X}_{L}),
\label{eqS:IR}
\end{equation}
where $I(\mathbf{S}_{L};\mathbf{X}_{L})$ is the mutual information
between the input trajectory $\mathbf{S}_{L}$ and the output
trajectory $\mathbf{X}_{L}$. The trajectories $\mathbf{S}_{L}=[S(t_1),
S(t_{2}), \dots, S(t_n)] $ and $\mathbf{X}_{L}=[X(t_1), X(t_{2}),
\dots, X(t_n)]$ are spin trajectories of duration $L=(n-1)\delta t$,
containing $n$ subsequent spin states $S$ and $X$, respectively, at
successive time points $t_i= (i-1) \delta t$, with $\delta t$ the
elementary timestep of the dynamics. Here, in Eq.  \ref{eqS:IR} the
limit $L\to \infty$ corresponds to $n\to \infty$ since we keep the
elementary time step $\delta t$ fixed. In this study the time is
measured in units of $\delta t$.

The mutual information $I(\mathbf{S}_L;\mathbf{X}_L)$ is given by the
entropies of the input and output:
\begin{equation}
I(\mathbf{S}_L;\mathbf{X}_L) =  H(\mathbf{S}_L) + H(\mathbf{X}_L) - H(\mathbf{S}_L,\mathbf{X}_L)
\end{equation}
where $H(\mathbf{S}_L)$ and $ H(\mathbf{X}_L)$ are the entropies of
the input and output trajectories, respectively, and
$H(\mathbf{S}_L,\mathbf{X}_L)$ their joint entropy.

\begin{figure}[t]
 \centering
 \begin{adjustbox}{center}
   \includegraphics[height=60mm,width=1.1\columnwidth]{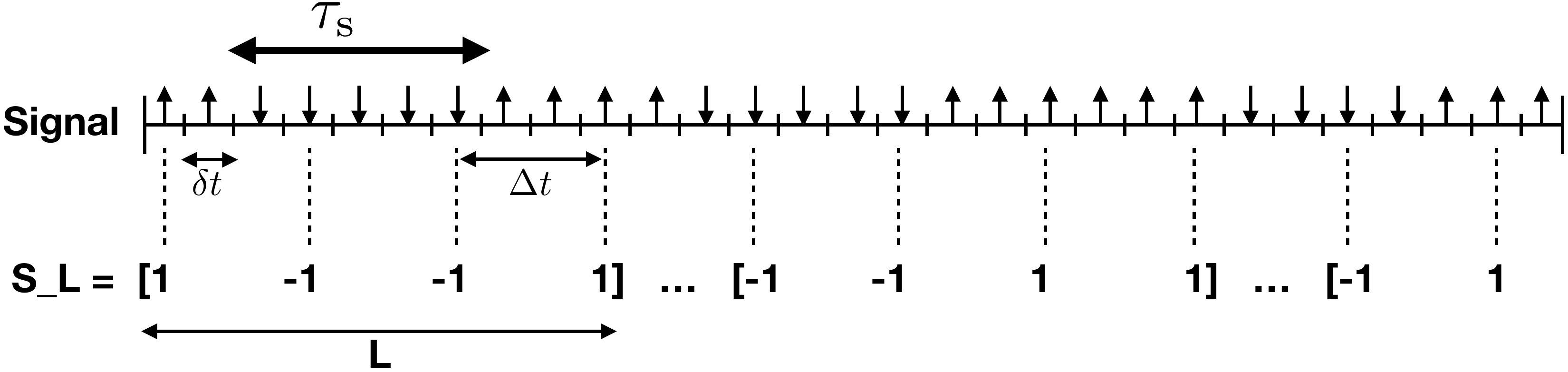}
 \end{adjustbox}
 \caption{Procedure for computing the entropies of the input and
   output trajectories, illustrated for a trajectory of the input signal. The
   signal produces a new spin state at each time step $\delta t$ and
   has a correlation time of $\tau_{\rm s}$. Samples of the trajectory
   $\textbf{S}_L(\Delta t)$ are collected by adding spin states to a
   trajectory each $\Delta t$ until the length of the trajectory has
   reached $L$. The trajectory is then saved as a single observation
   and (while continuing our simulation) we start collecting spin
   states for the next trajectory. Trajectories with the same sampling
   interval and length are thus collected in sequence and do not
   overlap with each other. During a given simulation trajectories of
   different lengths and sampling intervals are collected in
   parallel. The samples of the output and joint trajectories are
   collected in a similar manner.}
\label{fig:sampling}
\end{figure}

To estimate $I_{\text R}$, we compute the entropies  as a function of a sampling interval $\Delta t$ and the duration
of the trajectories $L = (n-1) \Delta t$, as illustrated in
Fig. \ref{fig:sampling} for the entropy $H({\bf S}_L,\Delta t)$ of the
input signal.
The quantity of interest, the information transmission rate $I_R$ is
then given by
\begin{equation}
I_{\text{R}} (\Delta t\to \delta t)= \frac{1}{L} \lim_{L \rightarrow
  \infty}I(\mathbf{S}_L; \mathbf{X}_L,\Delta t),
\label{eq:IRDt}
\end{equation} 
where 
\begin{equation}
I(\mathbf{S}_L;\mathbf{X}_L,\Delta t) = H(\mathbf{S}_L, \Delta t) + H(\mathbf{X}_L,\Delta t) - H(\mathbf{S}_L,\mathbf{X}_L,\Delta t),
\label{eq:MI_Dt}
\end{equation}
to emphasize that the mutual information and the entropies of the
input and output depend on the trajectory length $L$ and the sampling
interval $\Delta t$. The entropy of the trajectories increases with
decreasing $\Delta t$, even when  the
sampling interval $\Delta t$ is smaller than the timescale of the input signal
$\tau_{\rm s}$ and the the response time of the system $\tau_{\rm
  r}$. Secondly, the entropies also increase with the trajectory length $L$. For
sufficiently long $L > \tau_{\rm s}, \tau_{\rm r}$,
$I(\mathbf{S}_L;\mathbf{X}_L,\Delta t)$ grows linearly with $L$ at
fixed $\Delta t$, which means that the information transmission rate
is given by the slope of the mutual information
$I(\mathbf{S}_L;\mathbf{X}_L,\Delta t)$ as a function of $L$.
The information transmission rate is therefore also given by
\begin{equation}
I_{\text{R}} (\Delta t\to \delta t)= \lim_{L \rightarrow \infty}\frac{I(\mathbf{S}_L; \mathbf{X}_L,\Delta
  t) - I(\mathbf{S}_{L-\Delta t}; \mathbf{X}_{L-\Delta t},\Delta
  t)}{\Delta t}.
\label{eq:I_R}
\end{equation} 
When $L$ is smaller than either $\tau_{\rm s}$ or $\tau_{\rm
  r}$, the autocorrelations in the input and output trajectories mean that the
trajectory probabilities do not factorise in the probabilities of
independent, uncorrelated shorter trajectories, causing potentially
non-linear growth of the mutual information at small $L$. For these
two reasons, it is necessary to estimate the information transmission
rate at a sampling interval $\Delta t = 1$ and for a trajectory length
$L$ that is larger than $\tau_{\text{s}}$ and
$\tau_{\text{r}}$.

Directly computing the mutual information at large enough $L$ and
$\Delta t= 1$ poses however a computational problem. There are $n = L
/ \Delta t + 1$ spin states in a single trajectory, giving us $K=2^n$
unique possible trajectories of the input and output signal. Since the
joint trajectory of input and output contains twice as many spin
states, $K=2^{2n}$ for the joint trajectory. Estimating the entropy of
a distribution where the number of states $K$ is larger than the
number of observations leads to a systematic bias in the estimation of
entropy, which can only be solved by ensuring that the number of
observations $N >> K$. For a given number of samples $N$ and a given
sampling interval $\Delta t$, this limits the  maximum possible trajectory
length $L$ for which we can
reliably estimate the mutual information $I(\mathbf{S}_L;
\mathbf{X}_L,\Delta t)$.

Here we will elaborate on how strong the systematic bias due to
undersampling is, and for which combinations of $L$ and $\Delta t$ we
can reliably estimate the entropy of a trajectory. Then we will show
how we can get a reliable estimate of the information transmission
rate at the elementary time step, by computing the information
transmission rate for larger sampling intervals $\Delta t$ and then
extrapolating this value to $\Delta t = 1$ (again in units of the elementary
time step $\delta t$). In the sections {\bf Sampling parameter constraints} and {\bf
  Pseudo code} we summarize our procedure and the requirements that
$L$ and $\Delta t$ need to satify. In the section {\bf Pseudo code} we
also describe how we compute the response time $\tau_r$. We now first give background information.

\subsection*{Estimation of the entropy of the input signal}
Because it is possible to analytically compute the entropy of the
input signal, the inference of the entropy of the input signal
provides a good test case for our procedure to estimate the
entropy. At each elementary time step, there is a chance $r=\delta
  t / \tau_{\rm s}$ that the spin flips:
\begin{equation}
P(S_t = 1 | S_{t-1} = -1) = P(S_t = - 1 | S_{t-1} = 1) = r
\end{equation}
where $S_t$ is the state of the spin at time $t$ (in units of $\delta t$). Similarly,
\begin{equation}
P(S_t = 1 | S_{t-1} = 1) = P(S_t = - 1 | S_{t-1} = -1) = (1-r).
\end{equation}
The signal is a Markovian process since the chance of a spin 
flip does not depend on the history of the trajectory. The entropy
 rate of this process is then given by:
\begin{equation}
h(\mathcal{S}) = H(S_{t} | S_{t-1}) = -r \log r - (1-r) \log (1-r),
\end{equation}
where we assume that the spin up and spin down states are 
equally likely. Using this entropy rate, the quantity of interest, the
true entropy of the input
signal, for $\Delta t \to \delta t$, is 
\begin{align}
H(\textbf{S}_L) &= H(S) + \frac{L}{\delta t} h(\mathcal{S}) \\
&= \log(2) - \frac{L}{\delta t} \big[ r \log r + (1-r) \log (1-r) \big].
\end{align}

\begin{figure}[t]
    \centering
    \includegraphics[height = 75 mm,width=0.7\textwidth]{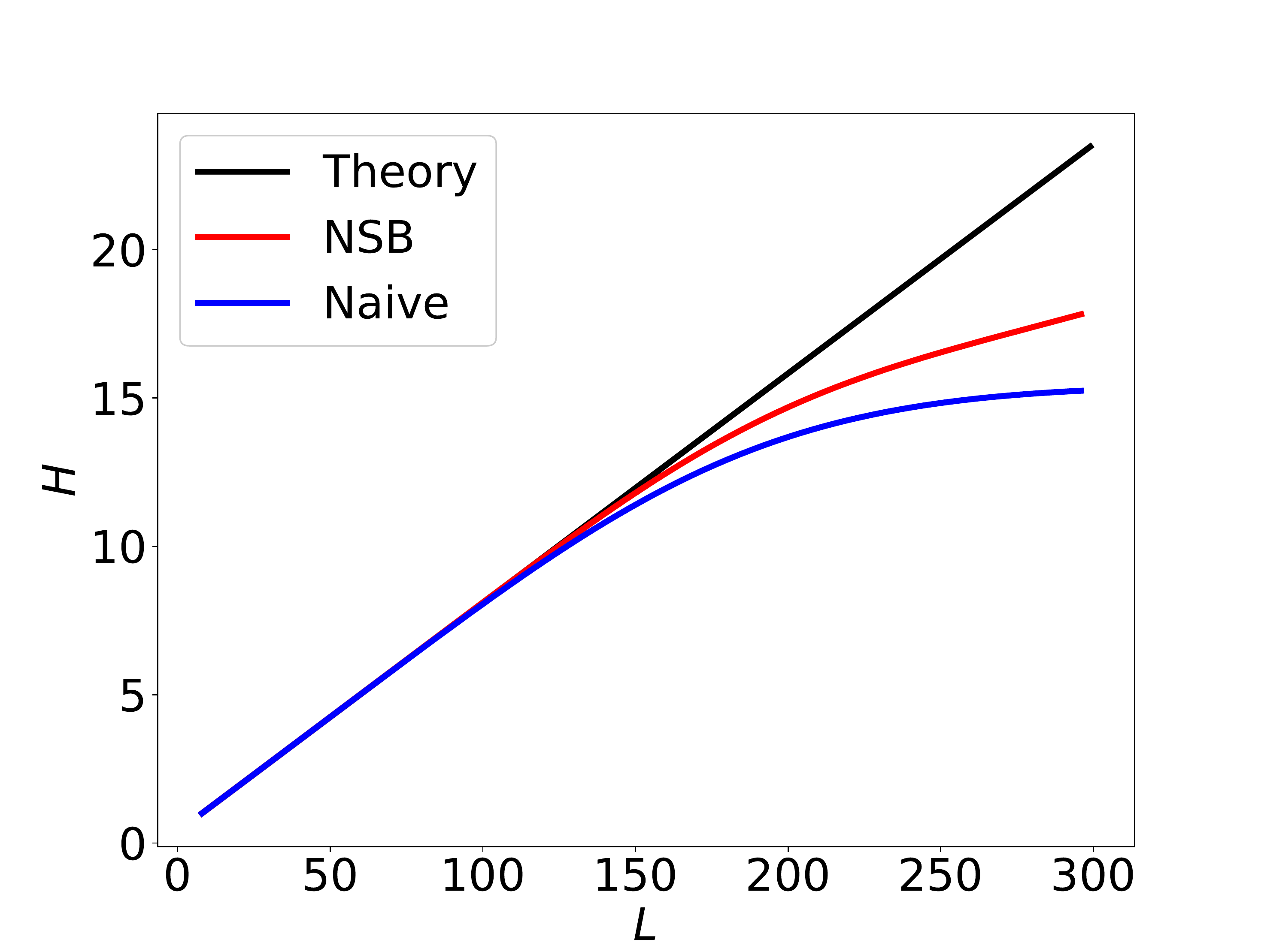}
    \caption{The NSB estimator outperforms the naive estimator when
      estimating the entropy of the input signal $H=H({\bf S}_L,\Delta
      t)$. Here we increase the
      trajectory length $L=(n-1)\Delta t$ by increasing the number of states $n$ in the
      trajectory, while keeping the sampling interval $\Delta t$
      constant at $\Delta t = 4$. Both estimators suffer from a
      systematic bias at larger $L$, where the entropy is
      underestimated. For small $L < 100$, there is sufficient
      sampling ($N_{\rm tot} = 5 \times 10^6$) and the estimators
      agree with the theoretically predicted value of the
      entropy. The correlation time of the input signal $\tau_{\rm s} = 40$.}
    \label{fig:Entropy_compare}
\end{figure} 

We now consider the effect of sampling the input trajectory at a
sampling interval $\Delta t$. The chance 
of a spin flip at the next time interval $\Delta t$ is
\begin{align}\label{eq:qdef}
P(S_t|S_{t-\Delta t}) = \begin{cases} \sum_{i=0,{\rm even}}^{\Delta t}
  (1-r)^{\Delta t -i} r^i \frac{\Delta t!}{i!(\Delta t - i)!} =
  q(\Delta t), \quad  &\text{for} \quad S_t = S_{t-
\Delta t} \\
\sum_{i=1,{\rm odd}}^{\Delta t} (1-r)^{\Delta t - i} r^i \frac{\Delta
  t!}{i!(\Delta t - i)!} = (1-q(\Delta t)) , \quad &\text{for} \quad
S_t = -S_{t-\Delta t},
\end{cases}
\end{align}
where we sum over each possible step where the spin could flip 
within a sampling interval $\Delta t$. The entropy rate is thus
\begin{align}
h(\mathcal{S},\Delta t) = H(S_t|S_{t-\Delta t}) &= - \sum_{S_{t-\Delta
  t}} P(S_{t-\Delta t}) \sum_{S_t} P(S_t|S_{t-\Delta t})\log
P(S_t|S_{t-\Delta t}) \\ 
&= q(\Delta t)\log (q(\Delta t)) + (1-q(\Delta t)) \log (1-q(\Delta t)),
\end{align}
such that the entropy is
\begin{equation}
H(\mathbf{S}_L,\Delta t) = \log(2) - \frac{L}{\Delta t}  \big[ q(\Delta t)\log (q(\Delta t)) + (1-q(\Delta t)) \log (1-q(\Delta t)) \big].
\end{equation} 
This expression reduces to Eq. 7, when $\Delta t /\delta t = 1$, as it
should. 

It is possible to compare this theoretical value of
$H(\mathbf{S}_L,\Delta t)$ with estimates of the entropy using
simulations of the input signal for a given sampling
interval. Different estimators have been proposed to estimate the
entropy \cite{Miller1974,Grassberger2008}. Here, we compare the naive
estimator, in which the probability of a specific trajectory is simply
given as $P(\mathbf{S}_L = \mathbf{s}_L) = N_{\mathbf{s}_L} / N_{\rm
  tot}$, where $ N_{\mathbf{s}_L} $ is the number of observations of
the trajectory $\textbf{s}_L$, and $N_{\rm tot}$ is the total number
of observations, to the estimator proposed by Nemenman \textit{et al.}
\cite{Nemenman2004}, called the NSB estimator. When the performance of
the two estimators are compared, we see, by comparing the
computational estimates to the theoretical value, that the NSB
estimator has overall a smaller error than the naive
estimator. However, both of these estimators suffer from a bias at
large $L$, where the entropy of the input signal is underestimated
because the number of states of the input trajectory, $K = 2^n$,
exceeds the number of observations. We have chosen the NSB estimator,
but also have developed a procedure to estimate the information
  transmission rate without undersampling.

\subsection*{For sufficient sampling, we can reliably estimate the mutual information}
\begin{figure}[t]
 \centering
 \begin{adjustbox}{center}
   \includegraphics[height=80mm,width=0.75\columnwidth]{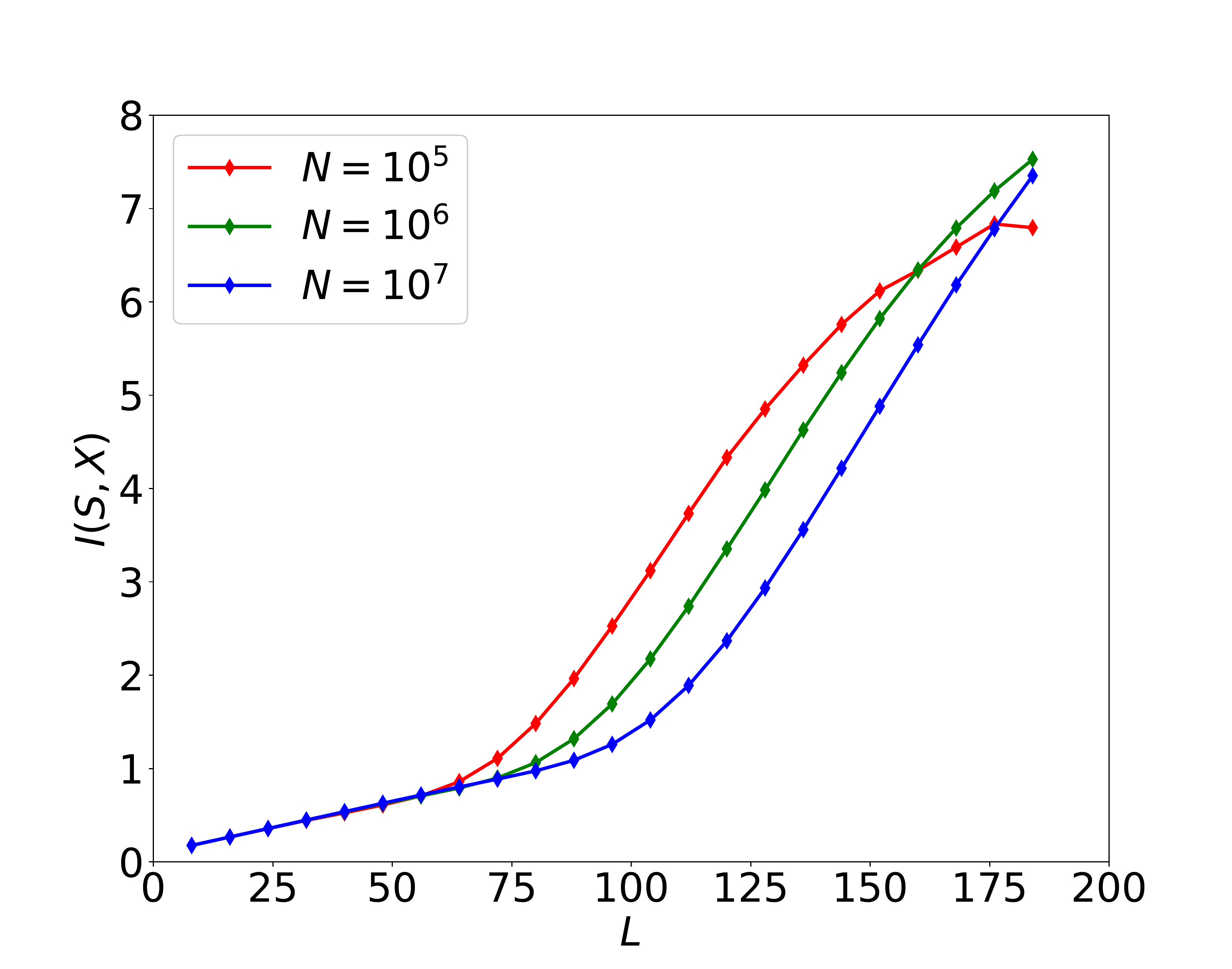}
 \end{adjustbox}
 \caption{The mutual information $I({\bf S}_L;{\bf X}_L,\Delta t)$ between the input spin and output
   spin as a function of the trajectory length $L=(n-1)\Delta t$,
   where we increase the length $L$ of the trajectories $\textbf{S}_L$
   and $\textbf{X}_L$ by increasing the number of spin states $n$ in
   the trajectory keeping the sampling interval $\Delta t$
   constant. Due to undersampling we can observe three regimes of our
   estimate of $I(\textbf{S}_L,\textbf{X}_L,\Delta t)$: Initially the entropies
   of the three trajectories are all correctly estimated such that the
   initial slope is the true information transmission rate. Then we
   underestimate only the joint entropy
   $H(\mathbf{S}_L,\mathbf{X}_L,\Delta t)$ (see Eq. \ref{eq:I_R}), which increases the mutual
   information. Finally, all three entropies are underestimated such
   that the mutual information again decreases. When we increase the
   number of observations $N$, we can elongate the length of the
   correctly estimated regime. The size of the system is $3 \times 3$
   spins at a temperature $T=2.4$, using a sampling interval of
   $\Delta t = 8$ and the correlation time of the input signal $\tau_{\rm s} = 25$.}
    \label{fig:Naive_MI}
\end{figure}

When we increase the length of the trajectories $L$ by increasing $n$
keeping $\Delta t$ constant, we can distinguish three regimes for the
mutual information $I({\bf S}_L; {\bf X}_L,\Delta t)$, see
Fig. \ref{fig:Naive_MI}: first, the mutual information increases with
a low, constant rate. Here, we can reliably estimate the entropies of
all three ensembles $\{\textbf{S}_L\}(\Delta t)$,
$\{\textbf{X}_L\}(\Delta t)$, and
$\{\textbf{S}_L,\textbf{X}_L\}(\Delta t)$ and the slope of the mutual
information equals the information transmission rate $I_{\rm R}(\Delta
  t)$ for this value of the sampling interval $\Delta t$.  Then as
  $L=(n-1)\Delta t$ is increased further (by raising $n$), the mutual
  information rises at a higher pace. In this regime, only the
entropy estimation of the joint trajectory
$\{\textbf{S}_L,\textbf{X}_L\}(\Delta t)$ suffers from
undersampling, causing $H({\bf S}_L;{\bf X}_L,\Delta t)$ to be
underestimated. The joint trajectory suffers from
undersampling first, because it contains twice the number of spin
states as compared to the input or output trajectories. Since the
entropy $H({\bf S}_L;{\bf X}_L,\Delta t)$ of the joint trajectory is
subtracted from the mutual information $I({\bf S}_L;{\bf X}_L,\Delta t)$, its underestimation
will cause the mutual information to be overestimated (see Eq.
\ref{eq:MI_Dt}). Finally, at larger values of $L$, the slope of the
mutual information decreases again. All three entropies are
now underestimated and the slope of the mutual information decreases. 
  Clearly, only in the initial linear regime, the
  information transmission rate can be reliably inferred from the
  slope of the mutual information $I({\bf S}_L;{\bf X}_L,\Delta t)$.

By increasing the number of observations, the initial regime is valid
for a larger range of trajectory lengths $L$. By increasing the number
of observations with a factor of $10^2$, the correct regime is
elongated with approximately $n\approx 5$ spin states
in the trajectories. Additionally, the collapse of all three lines in
the initial regime gives us confidence that we can reliably estimate
the mutual information when the trajectory does not contain too many
spin states. When the estimate of the mutual information does not
change when we repeat the simulation with more observations, then we
can be confident of our estimate of the mutual information: as we saw
in Fig. \ref{fig:Entropy_compare}, the NSB estimator does not have
any bias when there is sufficient sampling.

From inspection of Fig. \ref{fig:Naive_MI}, we see that for $N
= 10^7$ observations, the mutual information
$I(\textbf{S}_L;\textbf{X}_L,\Delta t)$ stays in the initial, correct, linear
regime up to $L \approx 72$, which corresponds to $n \approx L /
\Delta t + 1 = 72 / 8 + 1 \approx 10$ spin states in the trajectory, corresponding
to a state space of $K = 2^{2n} \approx 10^6$ for the join trajectory. For the
results of the main text, we have used $N = 4 * 10^7$ for $n = 9$ spin
states in the input and output trajectories. When increasing or decreasing
the number of spin states $n$, we adjusted the number of observations $N$
according to the change in the size of the state space $K$.
Using these parameters, we have $N >> K$ and there is a vanishingly
small error on the estimate of the mutual information.

\subsection*{The information transmission rate increases for a smaller sampling interval}
To reliably estimate the information transmission rate, it is
necessary to compute the rate at a sufficiently long trajectory length
$L$, which should be longer than the longest timescale in the system,
$L > \tau_{\rm s}, \tau_{\rm r}$, where $\tau_{\rm s}$ is the input
timescale and $\tau_{\rm r}$ the response time of the system. Yet,
  the number of spin states in the trajectories, $n$, cannot be too
  large because this will create a sampling problem, as discussed
  above. We thus need to increase the sampling interval $\Delta t$
  beyond $\delta t$.
However, for a given overall trajectory length $L=(n-1) \Delta t$, the
mutual information $I(\textbf{S}_L;\textbf{X}_L,\Delta t)$ depends on
$\Delta t$ while we would like to obtain the limit $\Delta t \to
\delta t$,
which is the elementary time step of the Glauber dynamics. 

Figure \ref{fig:IR_dt} illustrates how the information transmission
rate $I_{\rm R}(\Delta t)$ increases for decreasing sampling interval
$\Delta t$. The 6 black points show the computed information
transmission rate $I_{\rm R}(\Delta t)$ for 6 different values of $\Delta
t$. For these 6 values of $\Delta t$, the number of observations $N =
5 \times 10^7$ and the number of spins states in the trajectory $n=5$
is kept constant, such that $N \gg K$, where $K$ is the number of
unique possible trajectories (see previous section), and we can
reliably estimate the entropies of the trajectories. Because $n$ is
constant for these 6 black points, the trajectory length
$L=(n-1)\Delta t$ decreases for smaller $\Delta t$. Yet, the trajectories
  remain long enough, meaning that $L > \tau_{\rm s}, \tau_{\rm r}$
  (at the temperature $T = 2.45$ of the simulations, the response time
  $\tau_{\rm r} = 51$ and the correlation time of the input signal
  $\tau_{\rm s} = 63$). The mutual information $I({\bf S}_L;{\bf
    X}_L,\Delta t)$ thus increases linearly with $L$. From the
  slope of $I({\bf S}_L;{\bf X}_L,\Delta t)$ as a function of $L$, i.e.
  from Eq. \ref{eq:I_R}, we
  can therefore reliably estimate the information transmission rate
  $I_{\rm R}(\Delta t)$ for each value of $\Delta t$

To get the quantity of interest, the information transmission rate
$I_{\rm R}(\Delta t \to 1)$,  we fit a quadratic function to the estimates 
of $I_{\rm R}(\Delta t)$ at the 6 values of $\Delta t$ corresponding
to the black points. This function is the black dashed line. This
function is then extrapolated
to $\Delta t = 1$ to retrieve the value of the information
transmission rate at the elementary time step of the Glauber dynamics
(see the extrapolated black dashed line). In the case of
Fig. \ref{fig:IR_dt}, we find a value of {$I_{\rm R}(\Delta t = 1) =
  0.0048$.

 \begin{figure}[t]
 \centering
 \begin{adjustbox}{center}
   \includegraphics[height=85mm,width=0.8\columnwidth]{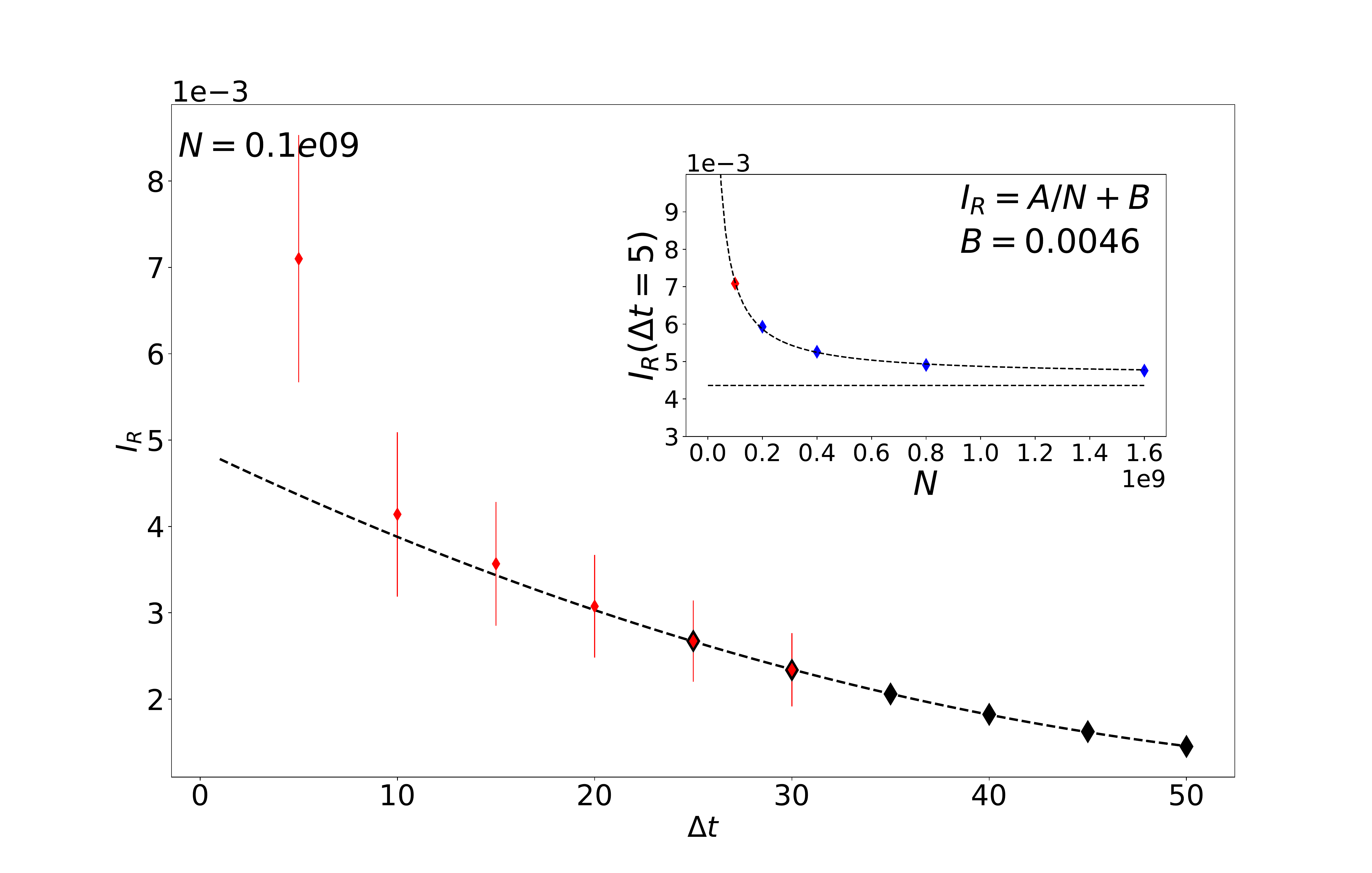}
 \end{adjustbox}
 \caption{The information transmission rate $I_{\text{R}}$ increases
   for decreasing sampling interval $\Delta t$. The 6 black points
     correspond to
   computations of $I_{\rm R}(\Delta t)$ at 6 values of $\Delta t$, obtained using
   $N=5 \times 10^7$ observations and $n =5$ spin states in a
   trajectory. $I_{\text{R}}(\Delta t =1)$ is estimated by
   extrapolating these measurements using a quadratic fit (dashed
   line). In order to verify this procedure, we have recomputed
   $I_{\text{R}}$ at smaller $\Delta t$ red
     points), where the number of observations $N = 10^8$. When
   decreasing the sampling interval $\Delta t$, we
   increase the number of spin states in the trajectory $n$ to make
   sure that the trajectory length $L > \tau_{\rm s}, \tau_{\rm r}$.
   At $\Delta t = 5$, the smallest sampling interval investigated, we
   use $n=15$. The computed $I_{\rm R}$ at the smaller sampling
   intervals (the red points)
   is larger than the extrapolated values (given by the dashed black
   line). This is due to undersampling, as illustrated in the inset
   for $\Delta t = 5$: by
   increasing the number of observations $N$, the estimated value
   decreases to our extrapolated estimate, the horizontal line in the
   inset. The size of the system is $5 \times 5$ spins, at a
   temperature of $T = 2.45$ and distance $d = 1$. The time scales are
   $\tau_{\rm r} = 51$ and $\tau_{\rm s} = 63$.}
    \label{fig:IR_dt}
\end{figure}

  In order to verify this scheme, we have recomputed the
    information transmission rate $I_{\rm R}(\Delta t)$ for a number
    of extrapolated values of $\Delta t$; these are the red
    points. For these smaller sampling intervals $\Delta t$ we have
  increased the number of spin states $n$ in the trajectories to
  ensure that the trajectory length remains long enough, ie. $L >
  \tau_{\rm s}, \tau_{\rm r}$. At the smallest sampling interval,
  $\Delta t = 5$, we used $n = 16$ such that $L = 75$.  For these
  values of $\Delta t$ corresponding to the red
  points, the number of samples is increased to $N=10^8$ and the
  errorbar is estimated using the NSB method. The figure illustrates
  that at small $\Delta t$ the computed values (red points)
  overestimate the information transmission rate as compared to the
  extrapolated values, given by the black dashed line. In the inset,
  we see however that this overestimation is due to undersampling: by
  increasing the number of observations $N$, the computed information
  transmission rate decreases to approach the extrapolated value,
  given by the horizontal line. The information transmission rate is
  inversely proportional to the number of observations, as is shown by
  the fit in the dashed line in the inset. Comparing the value that
  $I_{\rm R}$ decays to, $B = 0.0046$, with the extrapolated value
  from the main panel, $I_{\rm R}(\Delta t = 5) \approx 0.0044$,
  shows that our extrapolation procedure gives a reliable estimate of
  the information transmission rate for smaller $\Delta t$ values. We
  also note that this figure underscores the observation of Fig. 1,
  namely that even the NSB method suffers from undersampling.

 \begin{figure}[t]
 \centering
 \begin{adjustbox}{center}
   \includegraphics[height=70mm,width=0.85\columnwidth]{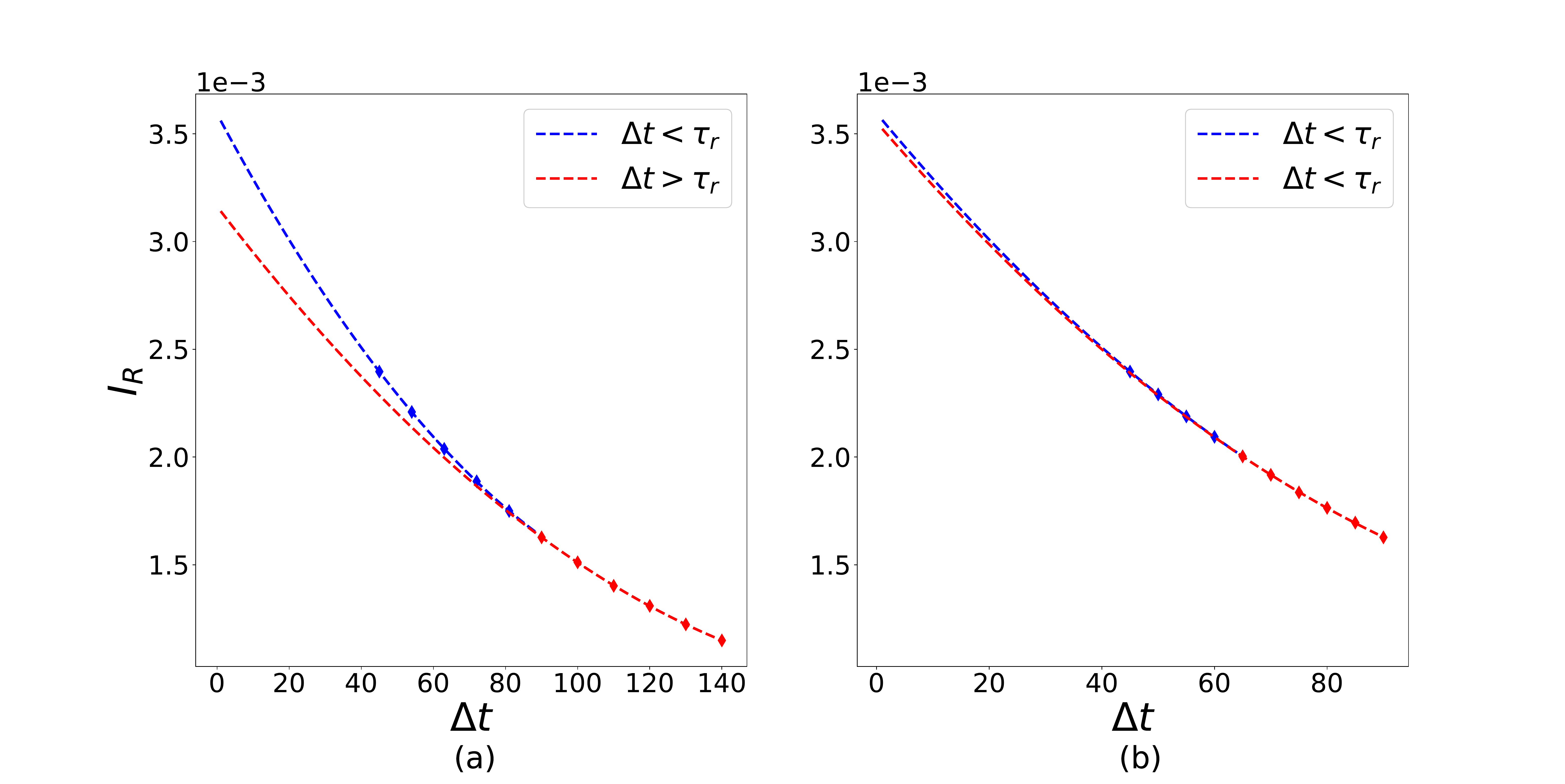}
 \end{adjustbox}
 \caption{The sampling intervals $\Delta t$ must be smaller than the
   correlation time $\tau_{\rm s}$ of the signal and the
   response time $\tau_{\rm r}$ of the system. Both panels show the information
   transmission rate $I_{\rm R}(\Delta t)$ in the same system of $5
   \times 5$ spins at temperature $T=2.3$, with timescales $\tau_{\rm
     r} = 88$ and $\tau_{\rm s} = 300$ and $n=9$ spin states in the
   trajectory. The number of observations $N = 10^7$ such that $N \gg
   K$. In panel a, the red dots correspond to a scheme in which the
   sampling intervals $\Delta t > \tau_{\rm r}$. In contrast, the blue
   dots correspond to a scheme in which $\Delta t < \tau_{\rm
     r}$. Clearly, the extrapolation in the red scheme underestimates
   the value of $I_{\rm R}(\Delta t = 1)$, because
   the sampling interval $\Delta t$ needs to be shorter than
   $\tau_{\rm r}$ and $\tau_{\rm s}$. This is also
   supported by panel b, in which {\em both} the red and blue schemes use
   sampling intervals $\Delta t$ that are all smaller than the response
   time. Both schemes result in essentially the same value of $I_{\rm
     R}(\Delta t=1)$, even though the extrapolation is based on
   different values of $\Delta t$.}
\label{fig:dt_vs_tc}
\end{figure}

\subsection*{The sampling interval must be smaller that the input
  correlation time and the response time of the system}
Above we discussed that the trajectory length $L=(n-1)\Delta t$ must
be larger than $\tau_{\rm s},\tau_{\rm r}$ and that $n$ cannot be too
large because of undersampling. Moreover, we described how the
information transmission rate of interest,
i.e. $I_{\rm R}(\Delta t)$ at $\Delta t =1$, can be obtained by extrapolating $I_{\rm R}
(\Delta t)$ computed for large $\Delta t$ to $\Delta t \to 1$ (see
Fig.  \ref{fig:IR_dt}). However, how do we choose the sampling
interval $\Delta t$ for which we compute the information transmission rate?  As
mentioned, it is necessary that $L = (n-1)\Delta t >
\tau_{\text{s}},\tau_{\text{r}}$; given that the maximum value of $n$
that allows for good sampling in reasonable CPU time is finite
$(n\approx 10)$, $\Delta t$ needs to be large enough. However, it is
not possible to indefinitely increase the sampling interval: $\Delta
t$ must be smaller than the correlation time of the input signal and the 
response time of the system. This is illustrated in
Fig. \ref{fig:dt_vs_tc}. Both panels correspond to a $5 \times 5$ spin
system at a temperature $T = 2.3$, in which the response
time $\tau_{\rm r} = 88$ and the correlation time of the input signal
$\tau_{\rm s} = 300$.  While keeping the number of spin states constant,
 $n=9$, $\Delta t$
and $L=(n-1)\Delta t$ are varied, in both panels. The two panels
differ in which range of $\Delta t$ values is used to extrapolate to
$\Delta t=1$. In panel a, the red dots and the red dashed line
correspond to a scheme in which the extrapolation is based on $\Delta
t$ values that are larger than the response time $\tau_{\rm r}$
  of the system}. In contrast, the blue dots and blue dashed line
correspond to a scheme in which the extrapolation is based on $\Delta
t$ values that are all smaller than $\tau_{\rm r}$. Clearly, the
extrapolation of the red scheme, based on $\Delta t$ values larger
than $\tau_{\rm r}$, severely underestimates the extrapolated value of
$I_{\rm R}$. We thus need to use $\Delta t$ values that are shorter
than $\tau_{\rm s},\tau_{\rm r}$. This is further supported by panel
b. In this panel, two extrapolation schemes are shown, which differ in
the values of $\Delta t$ used for the extrapolation. In contrast to
panel a, however, {\em both} of these schemes use $\Delta t$ values
that are all smaller than $\tau_{\rm s},\tau_{\rm r}$. Clearly, both
schemes give essentially the same extrapolated value of $I_{\rm R}$,
even though the extrapolation is based on different values of $\Delta
t$. 

\subsection*{Sampling parameter requirements} In summary, the
parameters of the sampling procedure must satisfy the following constraints: 
\begin{enumerate}
\item $\Delta t$ must be smaller than $\tau_{\rm s},
\tau_{\rm r}$
\item yet $L=(n-1) \Delta t$ must be larger than $\tau_{\rm
  s}, \tau_{\rm r}$. 
\item $N$ must be larger than $2^{2n}$ so that
undersampling does not occur. 
\end{enumerate}
When these three criteria are met, the
extrapolation procedure illustrated in Figs. \ref{fig:IR_dt} and
\ref{fig:dt_vs_tc} yields a reliable estimate of $I_{\rm R} (\Delta t=1)$.

\subsection*{Pseudo-code}
\begin{itemize}
\item{We have computed for each temperature $T$ the correlation time
    $\tau_{\text{c}}$ from the decay of the two-point time correlation
    function $\langle S(0) X(t)\rangle$, which serves as a measure for
    the response time of the system, $\tau_{\text{r}}$; in fact, simulations reveal that this
response time is similar to the timescale over which the total
magnetisation of the system relaxes to zero when the input spin, which
had been held fixed, is allowed to thermally equilibrate, indicating
that the driving of the system via the flips of the input spin keeps
the system in the linear-response regime \cite{Chandler1987}. In the optimal
    systems that maximize information transmission, $\tau_{\rm r}$ is
    typically on the order of the correlation
    time $\tau_{\rm s}$ of the input signal.}

\item{For a given temperature $T$ and correlation time of the input
    signal $\tau_{\text{s}}$, we choose six sampling intervals $\Delta
    t < \tau_{\rm s},\tau_{\rm r}$.
}

\item{For each sampling interval $\Delta t$, we calculate
$I_{\text{R}}(\Delta t)$ and $I_{\text{inst}}$. The number of
observations $N$ and spin states $n$ in a trajectory depend on the
temperature. At low temperatures, the long response time allows
for relatively large sampling intervals such that we use $n=8$ and
$N = 10^7$, while still having trajectory lengths $L >
\tau_{\rm s}, \tau_{\rm r}$. At higher temperatures, the response
time becomes shorter, such that the maximum sampling interval
$\Delta t$ that we can use also decreases. In order to still
satisfy the condition on the trajectory lengths $L$, we increase
the number of spin states in a trajectory to $n=10$ at $T=2.7$,
while increasing the number of observations accordingly to $N = 16
\times 10^7$, since $N>K = 2^{2n}$ for estimating the joint entropy.}

\item{Samples of the trajectories $\{\textbf{S}_{L}\} (\Delta t)$,
    $\{\textbf{X}_L\} (\Delta t)$, 
and $\{\textbf{S}_L,\textbf{X}_L\}(\Delta t)$ with different sampling intervals are 
collected in parallel when simulating the driven Ising system. As illustrated 
in Fig. \ref{fig:sampling}, the signal produces a new spin state every time step
$\delta t$ with a correlation time of $\tau_{\rm s}$. Samples for the trajectory
$\{ \textbf{S}_L\}$, which is characterised by a number of spin states $n$ and the
sampling interval $\Delta t$, are collected by storing spin states at every
$\Delta t$ in a vector until it reaches length $L$. Trajectories with the same 
sampling interval are collected in sequence and do not overlap, while trajectories
with different sampling intervals are collected in parallel and do overlap with each other. 
Samples of the output and joint trajectories are collected similarly.}

\item{Using the 
Bayesian estimator of Nemenman \textit{et al.} \cite{Nemenman2004}, 
we estimate the entropies of the (joint) trajectories and compute 
the mutual information and information transmission rate according 
to Eqs. 1 and 2 as a function of the sampling interval.}

\item{ By extrapolating the information transmission rate to $\Delta t
    =1$, we get the value $I_{\text{R}}(\tau_{\text{s}},T)$ that is
    plotted in Figs. 3 and 4 of the main text.}

\end{itemize}

\clearpage

\subsection*{The information rate in a system of $10 \times 10$ spins}
\begin{figure}[t]
 \centering
 \begin{adjustbox}{center}
   \includegraphics[height=90mm,width=0.9\columnwidth]{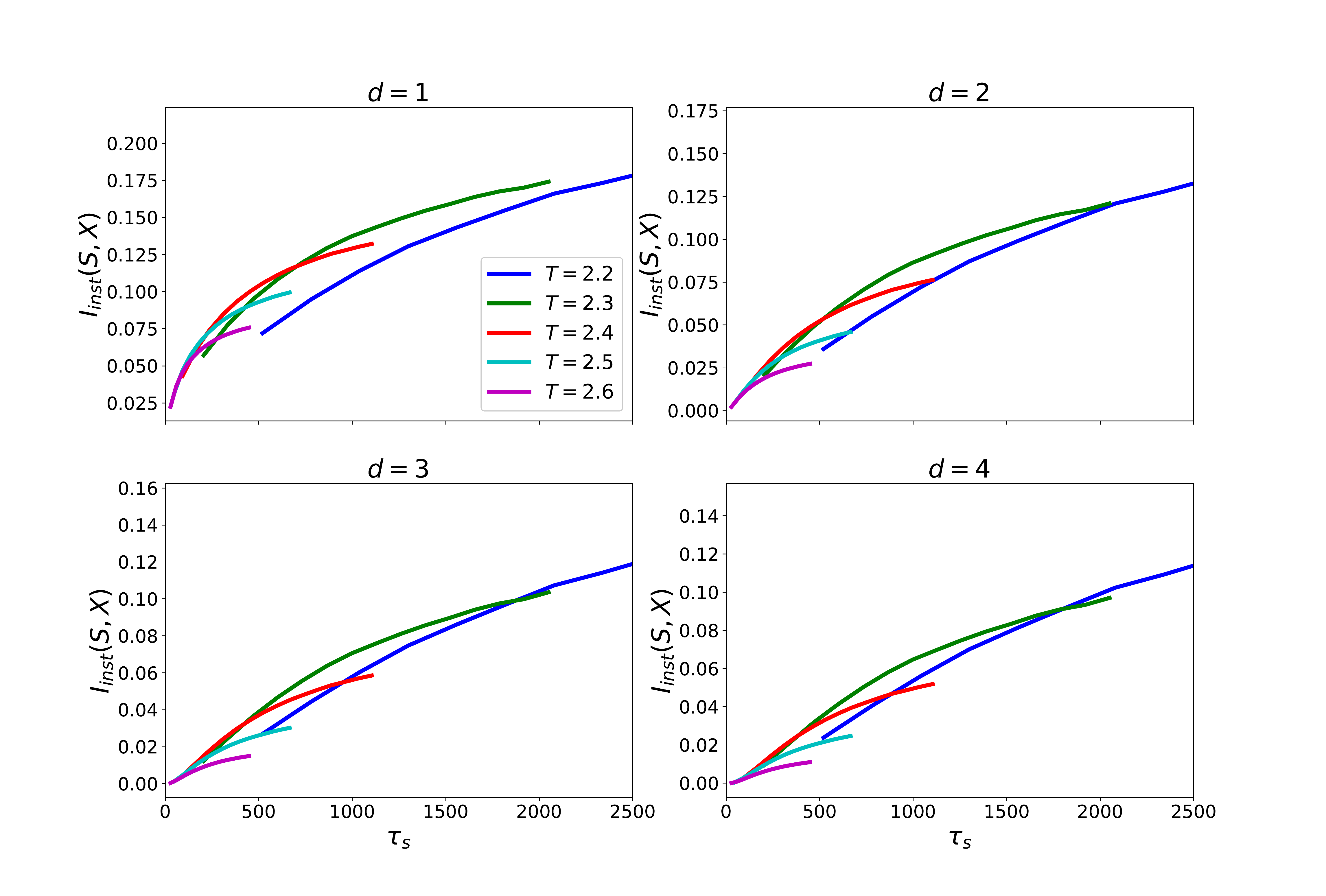}
 \end{adjustbox}
 \caption{The instantaneous mutual information $I_{\rm inst}(S;X)$ as
   a function of the input correlation time $\tau_{\rm s}$, for
   different temperatures $T$, and for different distances $d$ between
   the input and output spin, for a system of $10\times 10$
     spins. $I_{\rm inst}(S;X)$ increases monotonically with
   $\tau_{\rm s}$ until it reaches a plateau, which equals the static
   mutual information. This plateau increases with decreasing
   temperature because of the reduced thermal noise. At low
   temperatures, the plateau value is reached at considerably longer
   $\tau_{\rm s}$ than at high temperatures, reflecting the increased
   response times at low temperatures. Increasing the distance
   between the input and output spin, reduces the static mutual
   information at all temperatures. Note that also that for small
   $\tau_{\rm s}$ there exists an optimal temperature that maximizes
   $I_{\rm inst}$. It results from a trade-off between the necessity
   to respond fast and to respond reliably.}
\label{fig:10I_inst}
\end{figure}

\begin{figure}[t]
 \centering
 \begin{adjustbox}{center}
   \includegraphics[height=90mm,width=0.90\columnwidth]{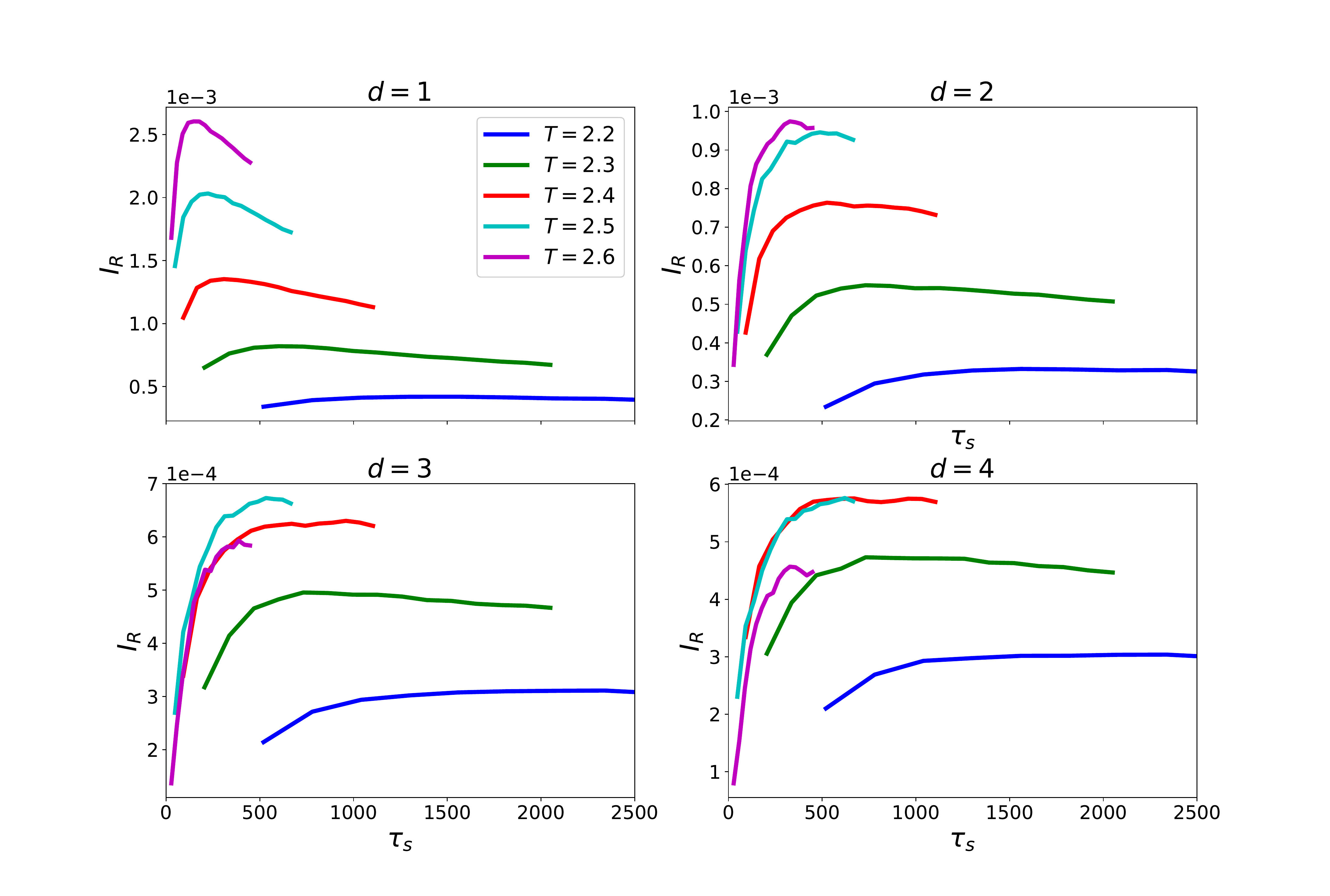}
 \end{adjustbox}
 \caption{The information transmission rate $I_{\rm R}$ as a function
   of the input correlation time $\tau_{\rm s}$ and temperature $T$
   for different distances $d$ between the input and output spin. For
   a given temperature, there exists an optimal $\tau_{\rm s}$ that
   optimizes $I_{\rm R}$, even though the curve is flattened at lower
   temperatures, where the system can only respond very
   slowly. Increasing $\tau_{\rm s}$ gives the system more time to
   respond to the input, which enhances the reliability of information
   transmission. Yet, increasing $\tau_{\rm s}$ also decreases the
   number of distinct input states that are transmitted per unit time,
   reducing the entropy of the input signal. This interplay gives rise
   to an optimal $\tau_{\rm s}$, at which $I_{\rm R}$ reaches it
   maximal value $I_{\rm R, max}$, which is plotted in Fig. 4a of the
   main text.}
\label{fig:10_IR}
\end{figure}

Similar to Figs. 3 and 4 of the main text, Figs. \ref{fig:10I_inst}
and \ref{fig:10_IR} show, respectively, the instantaneous mutual
information $I_{\rm inst}$ and the information transmission rate $I_{\rm
  R}$ as a function of the input signal correlation time
$\tau_{\text{s}}$ for a larger system, of $10 \times
10$ spins, and for different distances $d$. Because the effects of
criticality are stronger in the larger system, the differences in
response times $\tau_{\rm r}$ between different temperatures $T$
are larger. For this reason, we investigate a broader range of values
for $\tau_{\text{s}}$ in the two figures. Clearly, we find
qualitatively the same behaviour of $I_{\text{inst}}(S;X)$ and
$I_{\rm R}$ as in the system of $5 \times 5$ spins that is
investigated in the main text. Fig. 4a of the main text is
constructed by retrieving the maximum value of $I_{\rm R}$ at each
temperature in figure \ref{fig:10_IR}.

\clearpage

\subsection*{Scaling the system size}
\begin{figure}
 \centering
 \begin{adjustbox}{center}
   \includegraphics[height=90mm,width=\columnwidth]{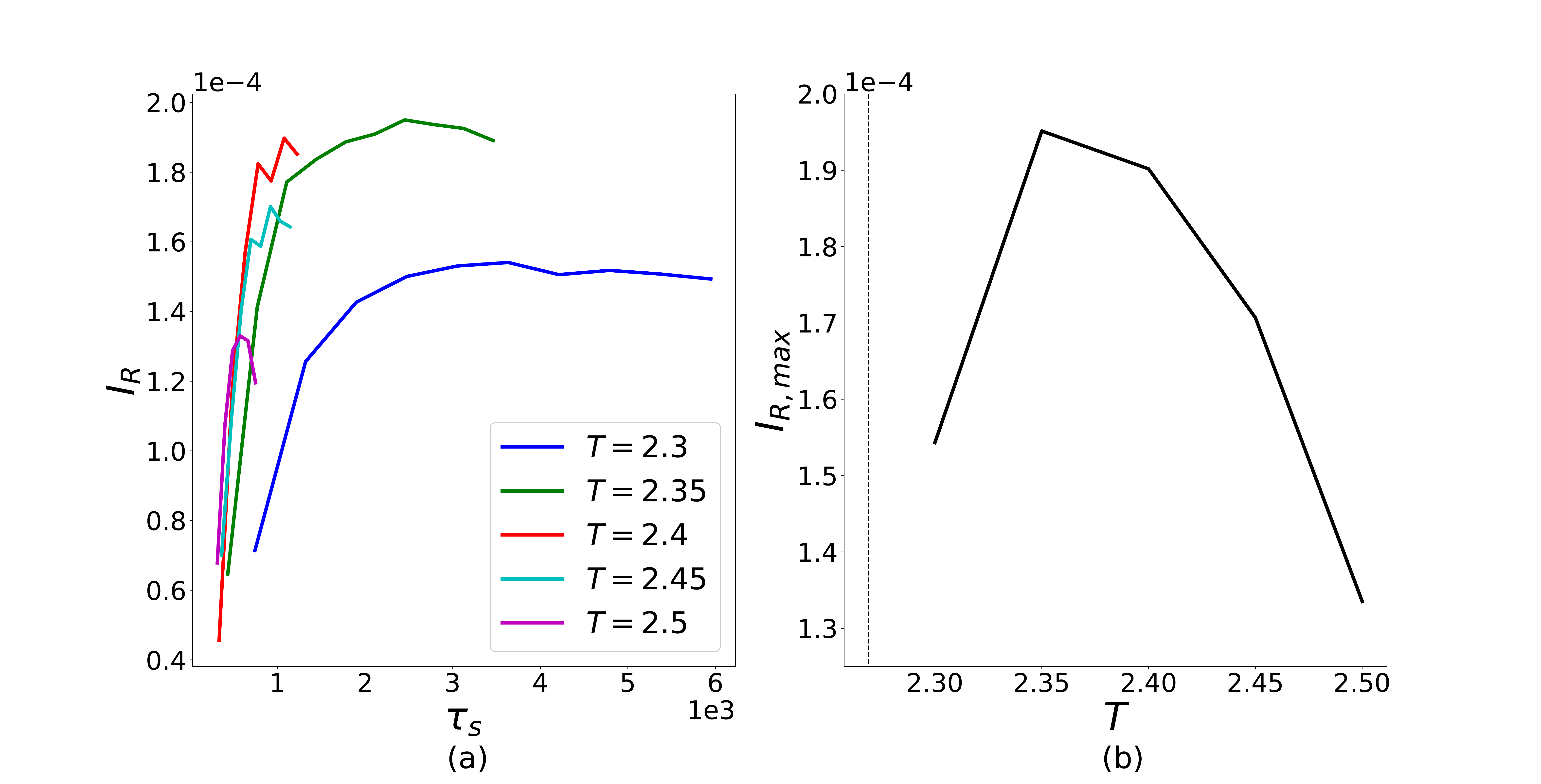}
\end{adjustbox}
 \caption{(a) The information transmission rate $I_{\rm R}$ as a function
   of the correlation time of the input signal $\tau_{\rm s}$ for
   different temperatures $T$ in a system of $15 \times 15$ spins at
   distance $d=6$. Due to the larger system size and distance between
   the input and output spin $d$, the sampling noise has increased
   significantly. However, there exists an optimal $\tau_{\rm s}$ that
   maximizes $I_{\rm R}$ for each temperature $T$. This maximal
   information transmission rate $I_{\rm R, max}$ is plotted as a
   function of temperature in panel b. Clearly, there is an optimal
   temperature that maximizes $I_{\rm R, max}$. }
\label{fig:15_IR}
\end{figure}

In order to investigate the behaviour of the information transmission
rate as the distance $d$ between the input and output signal increases
{\em together} with the system size ${\cal N}$, we have repeated the
same computations in a system of $15 \times 15$ spins at a distance
between input and output spin of $d = 6$. Fig. \ref{fig:15_IR}a shows
the information transmission rate $I_{\rm R}$ as a function of the
correlation time of the input signal $\tau_{\rm s}$ for a range of
temperatures.  While the sampling noise has increased due to the larger
distance between the input and output spin, it is clear that there is
an optimal temperature $T_{\rm opt}$ that maximizes the information
transmission rate. Panel b of Fig. \ref{fig:15_IR} shows this maximum
information transmission rate $I_{\rm R, max}$ as a function of
temperature. It is seen that there is an optimal temperature $T_{\rm
  opt}$ that globally maximizes the information transmission rate for
this system. By fitting a quadratic function to this plot, we estimate
the optimal temperature to be $T_{\rm opt} = 2.38$.

\begin{figure}
 \centering
 \begin{adjustbox}{center}
   \includegraphics[height=70mm,width=0.6\columnwidth]{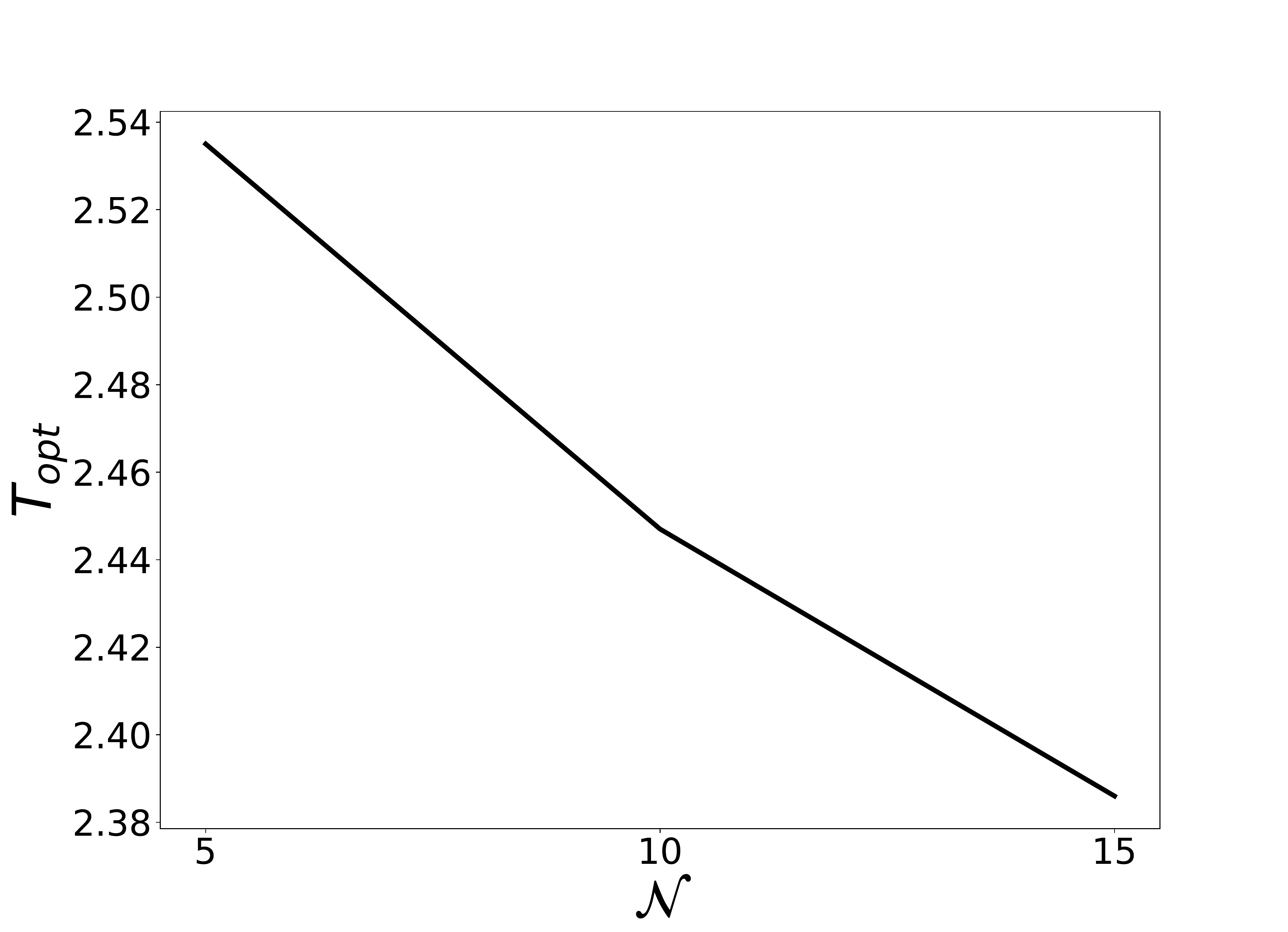}
\end{adjustbox}
\caption{The optimal temperature $T_{\rm opt}$ that maximizes the
  information transmission rate decreases as the distance $d$ is
  scaled {\em together} with the system size $\mathcal{N}$ from
  ($d=2,\mathcal{N} = 5$), to ($d=4, \mathcal{N} = 10$) and ($d =6,
  \mathcal{N} = 15$).}
\label{fig:Scaling}
\end{figure}

Fig. \ref{fig:Scaling} shows the optimal temperature as a function
  of system size ${\cal N}$, scaling the distance $d$ between input
  and output spin together with system size ${\cal N}$. The three points
  correspond to ($d=2, {\cal N}=5)$, ($d=3, {\cal N}=10$), and ($d=4,
  {\cal N}=15$). It is seen that the optimal temperature that
  maximizes the information transmission rate moves in the direction
  of the critical temperature as the system size is increased.

\end{document}